\documentstyle[amstex,pre,aps,multicol,epsf]{revtex}
\def\R{{\mathbb R}}  
 
\def\Z{{\mathbb Z}}
\newcommand{\bb}[1]{\begin{equation} #1 \end{equation}}
\newcommand{\e}{\mbox{ }}
\newcommand{\ba}[1]{\begin{eqnarray} #1 \end{eqnarray}}

\begin{document}

\title{Extremal-point Densities of Interface Fluctuations}
\author{Z. Toroczkai,$^{1,2}$ G. Korniss,$^3$ S. 
Das Sarma$^{1}$, and
R.K.P. Zia$^2$}
\address{$^1$Department of Physics, University of 
Maryland, College Park, 
MD 20742-4111 \\
$^2$Department of Physics, Virginia Polytechnic 
Institute and State 
University, Blacksburg, VA 24061-0435 \\
$^3$Supercomputer Computations Research Institute, 
Florida State 
University, Tallahassee, Florida 32306-4130 \\}

\date{\today}
\maketitle

\begin{abstract}
We introduce and investigate the stochastic dynamics of 
the density of local extrema (minima and maxima) of non-equilibrium
surface fluctuations. We give a number of exact, analytic
results for interface fluctuations described by  linear 
Langevin equations, and for on-lattice,
solid-on-solid surface growth models. We show that in spite
of the non-universal character of the quantities studied, their
behavior against the variation of the microscopic length scales  
can present generic features, characteristic to the macroscopic  
observables of the system. The quantities investigated here present  
us with tools that give an  entirely un-orthodox approach to the  
dynamics of surface morphologies:  a statistical analysis  from  
the short wavelength  end of  the Fourier decomposition spectrum.  
In addition to surface growth applications, our results can be used  
to solve the asymptotic scalability problem of massively parallel  
algorithms for discrete event simulations, which are extensively  
used in Monte-Carlo type simulations on  parallel architectures. 
\end{abstract}

\section{Introduction and Motivation}

The aim of statistical mechanics is to relate the macroscopic 
observables to the microscopic properties of the system. Before 
attempting any such derivation one always has to specify the 
spectrum of length-scales the analysis will comprise: while 
`macroscopic' is usually defined in a unique way by the 
every-day-life length scale, the `microscopic' is never so 
obvious, and the choice of the best lower-end scale is highly 
non-universal, it is system dependent, usually left to our 
physical `intuition', or it is set by the limitations of the 
experimental instrumentation. It is obvious that in order to 
derive the laws of the gaseous matter we do not need to employ
the physics of elementary particles,
it is enough to start from an effective  microscopic 
model (or Hamiltonian) on
the level of molecular  interactions.  Then starting from the equations of
motion on  the microscopic level and using a statistic and probabilistic
approach, the macroscale physics is derived. In this `long wavelength' 
approach most of the microscopic, or short wavelength 
information is usually redundant   and it is scaled away.            

Sometimes however, microscopic quantities are important and 
directly contribute to macroscopic observables, e.g.,  
the nearest-neighbor correlations in driven 
systems  determine the current,  
in model B the mobility, in kinetic Ising model the 
domain-wall velocity, in parallel computation the utilization
(efficiency) of conservative parallel algorithms, etc. Once a lower length
scale is set on which we can  define an effective microscopic dynamics, it
becomes meaningful to ask questions about local properties {\em at this
length scale},  e.g. nearest neighbor correlations, contour
distributions, extremal-point  densities, etc. These quantities are obviously
not universal, however their  {\em behavior} against the variation of the
length scales can   present qualitative and universal features.
Here we study the dynamics of macroscopically rough surfaces via 
investigating an intriguing miscroscopic quantity: the density of 
extrema (minima) and its finite size effects. We derive a 
number of analytical results about these quantities for a 
large class of non-equilibrium surface fluctuations described by 
linear Langevin equations, 
and solid-on-solid (SOS) lattice-growth models. Besides their
obvious relevance to surface physics our technique can be used
to show \cite{KTNR} the asymptotic scalability of conservative
massively parallel algorithms for discrete-event simulation, i.e., the fact 
that the {\em efficiency} of such computational schemes 
does not vanish with 
increasing the number of  processing elements, but it has a 
lower non-trivial 
bound. The solution of this problem is not only of practical 
importance from 
the point of view of parallel computing, but it has 
important consequences for
our understanding of systems with {\em asynchronous} parallel dynamics, 
in general. 
There are numerous dynamical systems both man-made, and 
found in the nature,
that contain a ``substantial amount'' of parallelism. For example, 

1) in wireless cellular communications the call arrivals and departures are 
happening in continuos time (Poisson arrivals), and the discrete events (call
arrivals) are {\em not} synchronized by a global clock. Nevertheless, calls 
initiated in cells substantially far from each other can be processed 
simultaneously by the parallel simulator {\em without} 
changing the poissonian nature of the underlying process. 
The problem of designing efficient dynamic channel allocation schemes 
for wireless networks is a very difficult one and 
currently it is done by modelling the  network as a system of   interacting
continuous time stochastic automata on parallel architectures  
 \cite{GLNW}.  
 
2) in  magnetic systems the discrete events are the spin flip attempts  
(e.g.,  Glauber dynamics for Ising systems). While traditional single spin  
flip dynamics may seem inherently serial, systems with short range interactions 
can be simulated in parallel: spins far from each other with {\em different}  
local simulated times can be updated simultaneously. 
Fast and efficient parallel Monte-Carlo 
algorithms are extremely welcome when studying metastable decay and hysteresis 
of kinetic Ising ferromagnets below their critical temperature, see \cite{KNR} 
and references therein. 
 
3) financial markets, and especially the stock market 
is an extremely dynamic, high connectivity network of relations, thousands of 
trades are being made asynchronously every minute.  
 
4) the brain. The human  
brain, in spite of its low weight of approx. 1kg, and volume of 
1400 $cm^3$, it  
contains about 100 billion  neurons, each neuron being connected 
through synapses to approximately 10,000 other neurons. 
The total number of synapses in a human brain is about 1000 trillion
($10^{15}$). The neurons of a single human brain placed
end-to-end would make a ``string'' of an enourmous lenght: 
250,000 miles \cite{http}. Assuming that each neuron of a single human cortex
can be in two states only (resting or acting), the total number of
different brain configurations would be ${2^{10}}^{11}$. 
According
to Carl Sagan, this number is greater than the total number of protons 
and electrons of the known universe, \cite{http}.
The brain does  an incredible amount of parallel computation: it
simultaneously  manages all of 
our body functions, we can talk and walk at the same time, etc.  
 
5) evolution of networks such as the internet, has  parallel dynamics:  the 
local network connectivity changes concurrently as many sites are attached 
or removed in different locations. As a matter of fact the physics of such 
dynamic networks is a currently heavily investigated and a rapidly emerging 
field  \cite{alb}.  
 
In order to present the basic ideas and notions in the simplest way, in the 
following we will restrict ourselves to one dimensional interfaces that have no 
overhangs.  The restriction on the overhangs may actually be lifted with 
a proper parametrization of the surface, a problem to which we will return  
briefly in the concluding section. The first visual impression when we look 
at a rough surface $h(x,t)$ is the extent of the fluctuations perpendicular to 
the substrate, in other words, the {\em  width} of the fluctuations. The width 
(or the rms of the height $h$ of the fluctuations) is probably the most 
extensively studied quantity in interface physics, due to the fact that its 
definition is simply quantifiable and therefore measurable:  
\begin{equation} 
w(L,t) = \sqrt{\;\overline{\left[h(x,t)\right]^2}- 
\left[\overline{h(x,t)}\right]^2 }\;, \label{width}  
\end{equation} 
where the overline denotes the average over the substrate. 
It is well-known that this quantity characterizes the long wavelength 
behavior of the fluctuations, the high frequency components being averaged out 
in (\ref{width}).  The short wavelength end of the spectrum has been ignored 
in the literature mainly because of its non-universal character, and also 
because it seemed to lack such a simple quantifiable definition as the width 
$w$.  
 
In the following we will present a quantity that is almost as simple and 
intuitive as the width $w$ but it characterizes the high frequency 
components of the fluctuations and it is  simply quantifiable. 
For illustrative purposes let us consider the classic Weierstrass function 
defined as the $M \to \infty$ limit of the smooth functions $W_M(a,b;x)$: 
\bb{ 
W(a,b;x) = \lim_{M \to \infty} W_M(a,b;x)=\lim_{M \to \infty}  
\sum_{m=0}^{M} a^{-m} \cos\left( b^m x\right) 
\;,\;\;\;a,b > 1 \label{weier} 
}\e 
Figure 1a shows the graph of $W_M$ at $a=2$ and $b=3$ (arbitrary values) 
for $M=0,1,2,3,4$ in the interval $x\in [0,4\pi]$. As one can see, by
increasing  $M$ we are
adding  more and more detail to the graph of the function on finer and finer 
length-scales. Thus $M$ plays the role of a regulator for the microscopic 
cut-off length which is $b^{-M}$, and for $M = \infty$ and $b > a$, the
function  becomes nowhere differentiable as it was shown by Hardy 
\cite{Hardy}. 
 
Comparing the graphs of $W_M$ for lower $M$ values with those for  
higher $M$-s we observe that the width effectively does not
change, however 
the curves  look qualitatively very different. This is obvious from
(\ref{weier}):  adding an extra term will not change the long wavelength
modes, but adds  a higher frequency component to the 
Fourier spectrum of the graph.     
We need to operationally define a quantifiable parameter which makes a 
distinction between a much `fuzzier' graph, such as  $M=4$ and 
a  smoother one, such as $M=1$.
The natural choice based on Figure 1a is the 
{\em number of local minima} (or 
extrema) in the graph of function. In Figure 1b we present the number of  
local minima $u_M$ vs.  
$M$ for two different values of $b$, $b=2.8$ and $b=1.8$, while 
keeping $a$ at the same value of $a=2$.  
For all $b$ values (not only for these two) the leading behaviour is 
exponential: $u_M \sim \lambda^M$.  The inset in Figure 1b shows the 
dependence of the rate $\lambda$ as a function of $b$ for fixed $a$. We 
observe that for $b>a$, $\lambda=b$, but below 
$b=a$ the dependence crosses 
over to another, seemingly   
linear function. For $b>a$ the amplitude of the 
extra added term in $W_{M+1}$ is large enough to 
prevent the cancellation of 
the newly appearing minima  by the drop in the 
local slope of $W_M$.  At $b 
\leq a$ the number of cancelled extrema starts to 
increase drastically with an 
exponential trend, leading to the crossover seen in 
the inset of Figure 2b. 
It has been shown that the fractal dimension of the Weierstrass
function for $b>a$ is given by $D_0 = \ln{b} / \ln{a}$  \cite{Hunt},
\cite{Hughes}. For $b \leq a$ the Weierstrass curve becomes
non-fractal with a dimension of $D_0=1$. By varying $b$ with respect
to $a$, we are crossing a fractal-smooth transition at $b=a$.
The very intriguing observation we just come across  is that
even though we are in the smooth regime ($b<a$) the density of minima
is still a {\em diverging} quantity (the $b=1.8$ curve in Figure 1b).
It is thus possible to have an infinite number of  `wrinkles' in the
Weierstrass function without having a diverging length, without having a
fractal in the classical sense. The transition from fractal to smooth, as
$b$ is lowered appears  as a non-analyticity in the divergence rate
of the curve's wrinkledness.
A rigorous analytic treatment of this problem seems to 
be highly non-trivial and 
we propose it as an open question to the interested reader. 

\begin{figure}
\begin{minipage}{3.2 in}\epsfxsize=3.1 in  
\epsfbox{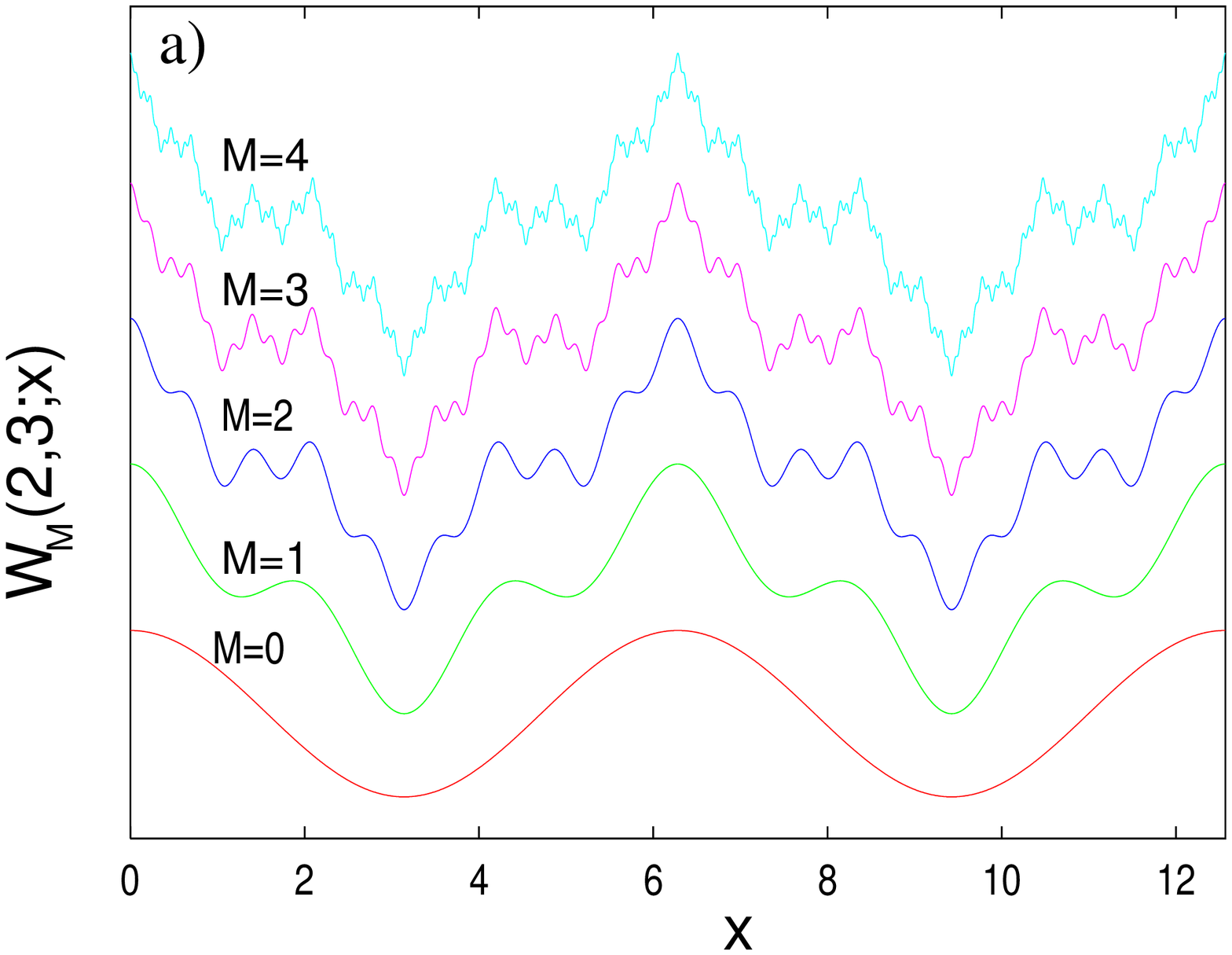} 
\end{minipage} 
\hspace*{-0.25cm} 
\begin{minipage}{3.5 in}\epsfxsize=3.4  
in \epsfbox{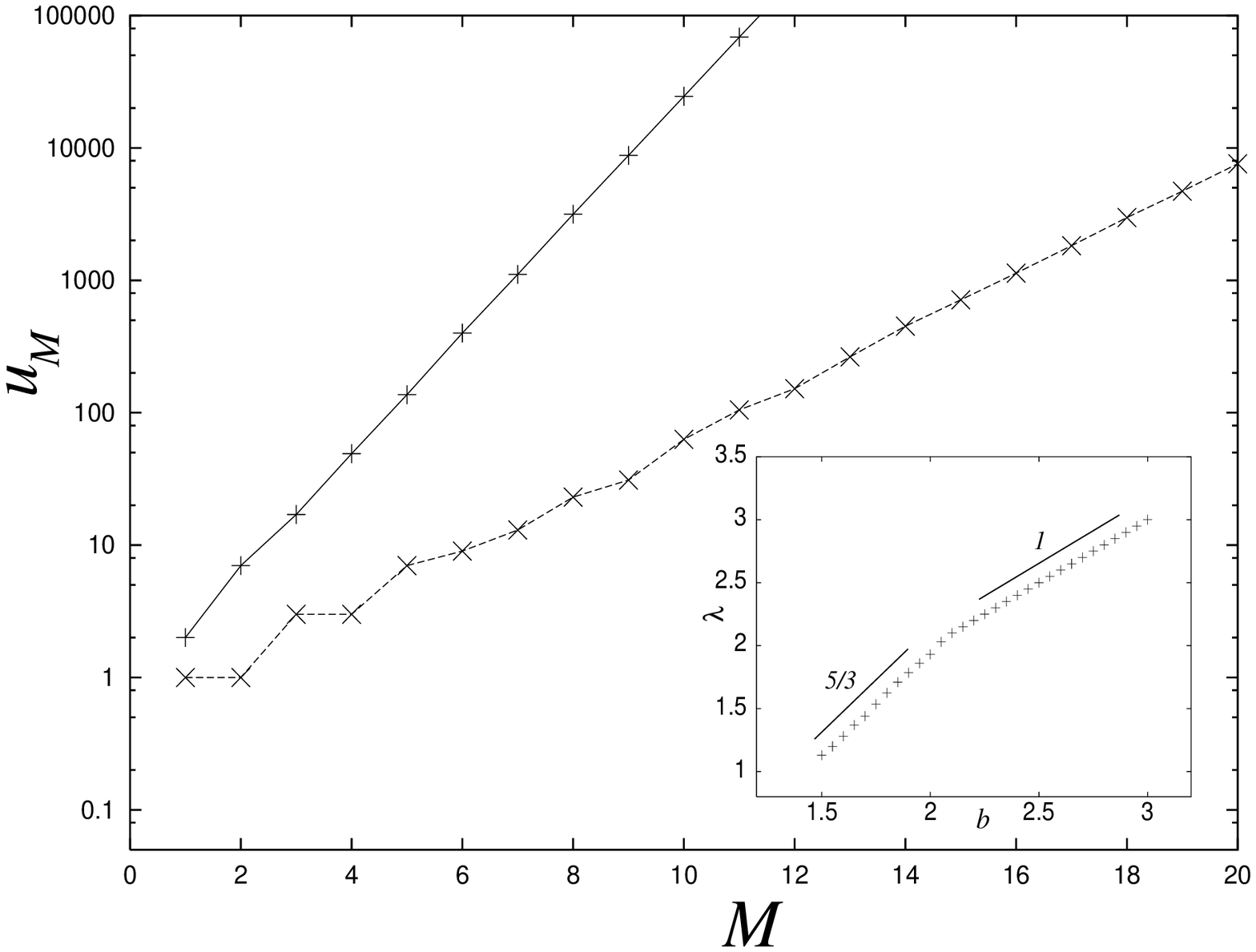} 
\end{minipage} 
\vspace*{0.5cm} 
\caption{a) the function $W_M(a,b;x)$ at $a=2$, 
$b=3$ and $M=0,1,2,3,4$. b) 
The scaling $u_M \sim \lambda^M$ of the number of local minima for 
$b=2.8$ (plusses) and $b=1.8$ (crosses). The inset shows $\lambda$ 
vs. $b$ while keeping $a$ constant, $a=2$.} 
\end{figure} 

 The simple example shown above suggests that there 
is novel and non-trivial 
physics lying behind the analysis of extremal 
point densities, and it  
gives extra information on the morphology of interfaces. 
Given an interface   $h(x,t)$, we 
propose a quantitative form that  characterizes the density of minima via a 
`partition-function' like expression, which is hardly more complex than
Eq. (\ref{width}) and  gives an alternate description of the surface
morphology:  
\bb{ 
u_q(L,[h]) = \frac{1}{L} \sum_{i} \left[ K(x_i)\right]^q\;,\;\;\;q>0\;,\;\;\; 
\mbox{$x_i$ are non-degenerate minima of $h$} \label{minima} 
}\e 
with $K(x_i)$ denoting the {\em curvature} of $h$ at the local 
(non-degenerate) 
minimum point $x_i$. The variable $q$ can be conceived as `inverse 
temperature'.  Obviously, for $q=0$ we obtain the number of local minima 
per unit interface length.  
The rigorous mathematical description and definitions lying at the basis of 
(\ref{minima}) is being presented in Section IV.  
The quantity in Eq. (\ref{minima}) 
is reminiscent to the partition function used in the definition of the  
thermodynamical formalism of one dimensional chaotic maps \cite{Ruelle} 
and 
also  to the definition of the dynamical or R\'enyi 
entropies of these chaotic 
maps. In that case, however the curvatures at the minima are replaced by 
cylinder intervals and/or the visiting probabilities of these cylinders. 
 
 We present a detailed analysis for the above quantity in case of  a 
large class of linear Langevin equations of type $\partial h / \partial t = 
-\nu (-\nabla^2 h)^{z/2}+\eta(x,t)$, where $\eta$ is a Gaussian noise term, 
and $z$ a positive real number. These Langevin equations are found to 
describe faithfully the fluctuations of monoatomic steps on various substrates, 
see for a review Ref. \cite{EDW}. One of the interesting conclusions 
we came to by studying the extremal-point densities on such equations 
is that depending on the value of $z$ the typical surface morphology 
can be fractal, or locally smooth, and the two
regimes are separated  by a critical $z$ value, $z_c$. In the fractal case, the
interface will have  infinitely many minima and cusps just as in the case of
the nondifferentiable  Weierstrass function (\ref{weier}), and the extremal
point densities  become infinite, or if the problem is discretized onto a
lattice with spacing  $a$, a power-law diverging behavior is observed as $a
\to 0^+$  for these densities. This sudden change of the `intrinsic roughness' 
may be conceived as a phase transition even in an experimental situation. 
Changing a parameter, such as the temperature, the {\em law} describing 
the fluctuations can change since the mechanism responsible for the 
fluctuations can change character as the temperature varies. 
For example, it has been recently shown using Scanning Tunneling 
Microscope (STM) measurements \cite{Giesen},
that the fluctuations of single atom layer steps on $Cu$ (111) below 
$T =300\mbox{ }^oC$ correspond to the perifery diffusion mechanism 
($z=4$), but above this temperature (such as $T = 500\mbox{ }^oC$ 
in their measurements) the mechanism is attachement-detachment 
where $z=2$, see also Ref. \cite{PT,Ted}.

The paper is organized as follows:  
In Sections II and III  we define and investigate on several well known
on-lattice models  the minimum point density and derive exact results in the
steady-state  ($t\to \infty$) including finite size effects. As a practical
application of  these on-lattice results, we briefly present in Subsection
III.B a lattice  surface-growth model which exactly describes the evolution of
the  simulated time-horizon for conservative massively parallel schemes in
parallel  computation, and solve a long-standing asymptotic scalability
question   for these update schemes.  In Section IV we lay down a more
rigorous  mathematical treatment for extremal point densities, and stochastic 
extremal point densities on the continuum, with a detailed derivation for 
a large class of linear Langevin equations (which are in fact the continuum 
counterparts of the discrete ones from Section II). The more rigorous treatment 
allows for an exact analytical evaluation not only in the steady-state, but 
for all times! We identify novel characteristic dimensions that separate 
regimes with divergent extremal point densities from convergent ones and 
which give a novel understanding of the short wavelength physics behind 
these kinetic roughening processes.

\section{Linear surface growth models on the lattice} 
 
In the present Section we focus on discrete, one 
dimensional models from the 
linear theory of kinetic roughening \cite{Barabasi,Krug}. Let us consider 
a one dimensional substrate consisting of $L$ lattice sites, with 
periodic boundary conditions. For simplicity the lattice constant is 
taken as unity, which clearly, represents 
the lower cut-off length for the effective equation of motion.  
For the moment let us study the discretized counterpart of the 
general Langevin equation that describes the linear theory 
of Molecular Beam Epitaxy (MBE) \cite{MBE,DT}: 
\begin{equation} 
\partial_{t} h_{i}(t) = \nu\nabla^2 h_{i}(t) - \kappa\nabla^4 h_{i}(t) 
+ \eta_{i}(t) \;, \label{h_evolution} 
\end{equation} 
where $\eta_{i}(t)$ is Gaussian white noise with  
\begin{equation} 
\langle\eta_{i}(t)\eta_{i}(t')\rangle = 2D\;\delta_{i,j}\;\delta(t-t') \;, 
\end{equation} 
and $\nabla^2$ is the discrete Laplacian operator, i.e.,  
$\nabla^2 f_j = f_{j+1}+f_{j-1} -2f_{j}$, 
applied to an arbitrary lattice function $f_j$. 
This equation arrises in MBE with both surface diffusion 
mechanism  (the 4th order, or curvature term)  and desorption mechanism 
(the 2nd order, or diffusive term)  present and it has been studied 
extensively by several authors \cite{MBE,Majaniemi}. Stability requires $\nu 
\geq 0$ and  $\kappa \geq  0$ (as a matter of fact, on the lattice is 
enough to have $\nu > 0$ and  $\kappa \geq -\nu /2$). Starting from a
completely flat initial   condition, the interface roughens until the
correlation length $\xi$ reaches  the size of the system $\xi \simeq L$, when
the roughening saturates over  into a steady-state regime. The process of
kinetic roughening is   controlled by the intrinsic length scale
\cite{Majaniemi},  $\sqrt{\kappa/\nu}$. Below this lengthscale the roughening
is dominated  by the surface diffusion or Mullins \cite{Mullins} term (the
4th order  operator) but above it is characterized by the evaporation piece
(the  diffusion) or Edwards-Wilkinson \cite{EW} term.  Since eq.
(\ref{h_evolution})  is translationaly invariant and linear in  $h$ it can be 
solved via the  discrete Fourier-transform:   
\begin{equation} \tilde{h}_k=\sum_{j=0}^{L-1} 
e^{-ikj}h_i \;,\;\;k=\frac{2\pi n}{L}\;,\;\;  n=0,1,2,\ldots,L-1 \;.  
\end{equation} 
Then eq. (\ref{h_evolution}) translates into  
\begin{equation} 
\partial_{t} \tilde{h}_{k}(t) = -\left[2\nu (1-\cos(k))  
 + 4\kappa (1-\cos(k))^{2}\right]\tilde{h}_{k}(t)+ \tilde{\eta}_{k}(t) \;, 
\label{h_k_evolution} 
\end{equation} 
with  
\bb{ 
\langle\tilde{\eta}_{k}(t)\tilde{\eta}_{k'}(t')\rangle =  
2D L \;\delta_{(k+k')\; \mbox{mod}\; 2\pi, 0}\;\delta(t-t') 
}\e 
Following the definition of the equal-time structure factor for $S(k,t)$, 
namely 
\bb{ 
S^{h}(k,t) L \delta_{(k+k')\; \mbox{mod}\; 2\pi, 0} \equiv  
\langle\tilde{h}_{k}(t)\tilde{h}_{k'}(t)\rangle \;, 
}\e 
one obtains for an initially flat interface:  
\begin{equation} 
S^{h}(k,t) = S^{h}(k) 
\left(1 - e^{-(4\nu (1-\cos(k)) + 8\kappa (1-\cos(k))^{2})t} 
\right)\;. 
\end{equation} 
In the above equation  
\begin{equation} 
S^{h}(k)\equiv \lim_{t\rightarrow\infty} S^{h}(k,t) = 
              \frac{D}{2\nu (1-\cos(k)) + 4\kappa (1-\cos(k))^{2}} 
\label{S_h} 
\end{equation} 
is the steady-state structure factor.  
 
For $\nu \neq 0$, and in the asymptotic scaling limit where 
$L\gg \sqrt{\kappa/\nu}$, the model belongs to the EW universality 
class and the roughening exponent is $\alpha=1/2$ (it is defined through the 
scaling $L^{2\alpha}$ 
of the interface width $\langle w_{L}^{2}(t)\rangle = 
(1/L)\langle\sum_{i=1}^{L}[h_{i}(t)-\overline{h}]^2 \rangle$ in the steady 
state). The presence of the curvature term does not change the universal 
scaling properties for the surface width,  and one finds the same exponents as 
for the pure EW ($\kappa=0$) case in eq.  (\ref{h_evolution}).  
For $\nu = 0$ the surface is purely curvature driven ($z=4$) and the model  
belongs to a different universality class where the steady-state width scales  
with a roughness exponent of $\alpha=3/2$. 
 
In the following we will be mostly interested in some local steady-state  
properties of the surface ${h_i}$. In particular, we want to find the density  
of local minima for the 
surface described by (\ref{h_evolution}). The operator which measures this 
quantity is \begin{equation} 
u = \frac{1}{L}\sum_{i=1}^{L} \Theta(h_{i-1}-h_{i}) \Theta(h_{i+1}-h_{i}) \;. 
\end{equation} 
This expression motivates the introduction of the local slopes, 
$\phi_i=h_{i+1} -h_{i}$. In this representation the operator for the density of 
local minima (for the original surface) is 
\begin{equation} 
u = \frac{1}{L}\sum_{i=1}^{L} \Theta(-\phi_{i-1}) \Theta(\phi_{i}) \;, 
\end{equation} 
and its steady-state average is  
$\langle u\rangle=\langle\Theta(-\phi_{i-1}) \Theta(\phi_{i})\rangle 
= \langle\Theta(-\phi_{1}) \Theta(\phi_{2})\rangle$, 
due to translational invariance. The average density of local minima is the  
same as the probability that at a randomly chosen site of the lattice 
the surface exhibits a local minimum. It is governed by the 
nearest-neighbor two-slope distribution, which is  
also Gaussian and fully determined by $\langle\phi_{1}^2\rangle= 
\langle\phi_{2}^2\rangle$ and $\langle\phi_{1}\phi_{2}\rangle$: 
\begin{equation} 
P^{nn}(\phi_1,\phi_2) \propto  
e^{-\frac{1}{2}\phi_j A^{\rm nn}_{jk}\phi_k} \;,\;\;\; j,k=1,2 \;, 
\end{equation} 
where 
\begin{equation} 
A^{\rm nn} = \left(\begin{array}{cc} 
\langle\phi_1^2\rangle         & \langle\phi_1\phi_2\rangle \\ 
\langle\phi_1\phi_2\rangle & \langle\phi_1^2\rangle 
\end{array} \right)^{-1} 
\end{equation} 
As we derive in Appendix A, the density of local minima only depends 
on the ratio $\langle\phi_1\phi_2\rangle/\langle\phi_1^2\rangle$: 
\begin{equation} 
\langle u\rangle = \frac{1}{2\pi}\arccos\left( 
   \frac{\langle\phi_1\phi_2\rangle}{\langle\phi_1^2\rangle}\right) \;.   
\label{min_dens} 
\end{equation} 
Finite-size effects of $\langle u\rangle$ are obviously carried through those
of   the correlations. First we find the steady-state structure factor 
for the slopes. Since $\tilde{\phi}_{k}=(1-e^{-ik})\tilde{h}_{k}$, we have  
$S^{\phi}(k)=2(1-\cos(k))S^{h}(k)$. Then from (\ref{S_h}) one obtains: 
\begin{equation} 
S^{\phi}(k) = \frac{D}{\nu + 2\kappa (1-\cos(k))}\;,\;\;\mbox{for}\;\; 
k \neq 0\;,\;\;\;\mbox{and}\;\;\;S^{\phi}(k)=0\;,\;\;\mbox{for}\;\; 
k=0\;. \label{S_phi} 
\end{equation} 
The former automatically follows 
from the $\sum_{i=1}^{L}\phi_{i}=0$ relation. Then we obtain the  
slope-slope correlations 
\begin{equation} 
C^{\phi}_{L}(l) \equiv \langle\phi_i\phi_{i+l}\rangle =  
\frac{1}{L}\sum_{n=1}^{L-1} e^{i \frac{2\pi n}{L} l} 
            S^{\phi}\left(\frac{2\pi n}{L}\right) \;.\label{C_L} 
\end{equation} 
With the help of Poisson summation formulas, in Appendix B we show a 
derivation for the exact spatial  correlation function, which yields 
\begin{equation} 
C^{\phi}_{L}(l) = \frac{D}{\nu + 2\kappa}\left\{ 
\frac{b^{|l|}}{\sqrt{1-a^2}} - \frac{1}{1-a}\frac{1}{L} + 
\frac{b^{L}}{1-b^{L}}\frac{b^{l}+b^{-l}}{\sqrt{1-a^2}}  
\right\} \;,\;\;\;|l|\leq L \;, \label{C_phi} 
\end{equation}  
with 
\begin{equation} 
a \equiv \frac{2\kappa}{\nu + 2\kappa}\;,\;\;\;\mbox{and}\;\;\; 
b \equiv \frac{1-\sqrt{1-a^2}}{a}\;.  
\end{equation} 
We have $|a| \leq 1$ and $b \leq 1$. 
The second term in the bracket in (\ref{C_phi}) gives a {\em uniform} power  
law correction, while the third one gives an exponential correction to the  
correlation function in the thermodynamic limit.  
For $\nu\neq 0$ and $L \rightarrow \infty$ one obtains 
\begin{equation} 
C^{\phi}_{\infty}(l) = \frac{D}{\nu + 2\kappa} \frac{b^{|l|}}{\sqrt{1-a^2}} = 
\frac{D}{\nu + 2\kappa} \frac{e^{-|l|/\xi^{\phi}_{\infty}}}{\sqrt{1-a^2}}  
\end{equation} 
where we define the correlation length of the slopes for an infinite system 
as: 
\begin{equation} 
\xi^{\phi}_{\infty} \equiv -\frac{1}{\ln(b)} \;. 
\end{equation} 
In the $\nu \rightarrow 0$ limit it becomes the 
intrinsic correlation length which diverges as $\nu^{-1/2}$: 
$ 
\xi^{\phi}_{\infty} \stackrel{\nu\rightarrow 0}{\simeq}  
\sqrt{\kappa / \nu}  
$ 
and 
\begin{equation} 
C^{\phi}_{\infty}(l)  \stackrel{\nu\rightarrow 0}{\simeq} 
\frac{D}{2\kappa}\left( \sqrt{\frac{\kappa}{\nu}} - |l| \right) \simeq 
\frac{D}{2\kappa}\left( \xi^{\phi}_{\infty} - |l| \right) \;. 
\label{C_inf_div} 
\end{equation} 
In this limit the slopes (separated by any finite distance) become highly  
correlated, and one may start to anticipate that the density of local minima 
will vanish for the original surface $\{h_i\}$.  
In the following two subsections we investigate the density of local minima 
and its finite-size effects for the Edwards-Wilkinson and the Mullins cases. 
 
\subsection{Density of local minima for Edwards-Wilkinson term dominated 
regime} 
 
To study the finite size effects for the local minimum density, we neglect  
the exponentially small correction in (\ref{C_phi}), so in the  
{\em asymptotic} limit, where  
$L\gg\xi^{\phi}_{\infty}$, $C^{\phi}_{L}(l)$ decays exponentially with  
{\em uniform} finite-size corrections: 
\begin{equation} 
C^{\phi}_{L}(l) \simeq \frac{D}{\nu + 2\kappa}\left\{ 
\frac{b^{|l|}}{\sqrt{1-a^2}} - \frac{1}{1-a}\frac{1}{L} 
\right\} \; 
\end{equation} 
This holds for the special case $\kappa=0$ as well, (in fact, there the  
exponential correction exatly vanishes) leaving 
\begin{equation} 
C^{\phi}_{L}(l) = \frac{D}{\nu}\left(\delta_{l,0}-\frac{1}{L}\right) \;. 
\end{equation} 
Now, emplying eq. (\ref{min_dens}), we can obtain the density of minima as: 
\begin{equation} 
\langle u\rangle_{L}= 
\frac{1}{2\pi}\arccos\left( \frac{C^{\phi}_{L}(1)}{C^{\phi}_{L}(0)} \right) 
\simeq \frac{1}{2\pi}\arccos(b) +  
\frac{1}{2\pi}\sqrt{\frac{1-b}{1+b}}\sqrt{\frac{1+a}{1-a}}\frac{1}{L} \;, 
\label{EW_min_dens} 
\end{equation} 
Again, for the $\kappa=0$ case one has a compact exact expression and the 
corresponding large $L$ behavior: 
\begin{equation} 
\langle u\rangle_{L} = \frac{1}{2\pi}\arccos\left( -\frac{1}{L-1}\right)  
\simeq \frac{1}{4} + \frac{1}{2\pi}\frac{1}{L} \;, \label{ezaz} 
\end{equation} 
which can also be obtained by taking the $\kappa\rightarrow0$ limit in 
(\ref{EW_min_dens}). 
To summarize, as long as $\nu\neq 0$, the model belongs to the EW universality 
class, and in the steady state, the density of local minima behaves as 
\begin{equation} 
\langle u\rangle_{L} \simeq \langle u\rangle_{\infty} + \frac{\rm const.}{L}\;, 
\label{EW2_min_dens} 
\end{equation} 
where $\langle u\rangle_{\infty}$ is the value of the density of local  
minima in the thermodynamic limit: 
\begin{equation} 
\langle u\rangle_{\infty} =\frac{1}{2\pi}\arccos(b) \;. 
\end{equation} 
Note that this quantity can be small, but does not vanish if  
$\nu$  is close but not equal to $0$. Further, the system exhibits the scaling  
(\ref{EW2_min_dens}) for asymptotically large systems, where  
$L\gg\xi^{\phi}_{\infty}$. It is important to see in detail how  
$\langle u\rangle_{\infty}$ behaves as $\nu \rightarrow 0$: 
\begin{equation} 
\langle u\rangle_{\infty} \stackrel{\nu\rightarrow 0}{\simeq} 
\frac{1}{2\pi}\arccos\left(1-\sqrt{2}\sqrt{1-a}\right) \simeq  
\frac{1}{2\pi}\arccos\left(1-\sqrt{\frac{\nu}{\kappa}}\right) \simeq 
\frac{1}{2\pi} \left(2 \sqrt{\frac{\nu}{\kappa}}\right)^{1/2} \simeq 
\frac{\sqrt{2}}{2\pi} \frac{1}{\sqrt{\xi^{\phi}_{\infty}}} \;. 
\label{u_inf_lim} 
\end{equation}  
Thus, the density of local minima for an {\em infinite} system vanishes as we 
approach the purely curvature driven ($\nu \rightarrow 0$) limit.  
Simply speaking, the local slopes become ``infinitely'' correlated, such that 
$C^{\phi}_{\infty}(l)$ diverges [according to  eq. (\ref{C_inf_div})], and the  
ratio  $C^{\phi}_{\infty}(l)/C^{\phi}_{\infty}(0)$ for any fixed $l$ tends to  
$1$. This is the physical picture behind the vanishing density of local minima.

\subsection{Density of local minima for the Mullins term dominated regime} 
 
Here we take the $\nu \rightarrow 0$ limit {\em first} and then study the 
finite  size effects in the purely curvature driven model. The slope 
correlations are finite for finite $L$ as can be seen from eq. (\ref{C_L}), 
since  the $n = 0$ term is not included in the sum! Thus, in the exact 
closed formula (\ref{C_phi}) a careful limiting procedure has to be  
taken which indeed yields the internal cancellation of the apparently 
divergent terms. Then one obtains the exact slope correlations 
for the $\nu=0$ case: 
\begin{equation} 
C^{\phi}_{L}(l) = \frac{D}{2\kappa} \left\{ 
\frac{L}{6}\left(1-\frac{1}{L^2}\right) -|l|\left(1-\frac{|l|}{L}\right) 
\right\} \; 
\end{equation} 
and for the local minimum density: 
\begin{equation} 
\langle u\rangle_{L} = \frac{1}{2\pi}\arccos\left(1-\frac{6}{L+1}\right) 
\simeq \frac{\sqrt{3}}{\pi}\frac{1}{\sqrt{L}}  \label{u4d}
\end{equation} 
It vanishes in the thermodynamic limit, and hence, one observes that 
the limits $\nu \rightarrow 0$ and  $L \rightarrow \infty$ are 
interchangable. For $\nu=0$, $L$ is directly  associated with the correlation 
length and we  can define $\xi^{\phi}_L \equiv L/6$. Then the correlations and 
the density of local minima takes the same scaling form as eqs. 
(\ref{C_inf_div})  and (\ref{u_inf_lim}): 
\begin{equation} 
C^{\phi}_{L}(l) \simeq  
\frac{D}{2\kappa}\left( \xi^{\phi}_{L} - |l| \right) \;, 
\end{equation} 
and 
\begin{equation} 
\langle u\rangle_{L}  
\simeq \frac{\sqrt{2}}{2\pi}\frac{1}{\sqrt{\xi^{\phi}_{L}}} \;. 
\end{equation}

\subsection{Scaling considerations for higher order equations} 
 
Let us now consider another equation but with a generalized 
relaxational term that includes the Edwards Wilkinson 
and the noisy Mullins equation as particluar cases: 
\begin{equation} 
\partial_{t} h_{i}(t) = -\nu \left( -\nabla^2\right)^{z/2} h  
+ \eta_{i}(t) \;. \label{kinetic} 
\end{equation} 
where $z$ is a positive real number (not necessarily integer). 
Other $z$ values of experimental interest are $z=1$, relaxation 
through plastic flow, \cite{Mullins,Krug}), and $z=3$ terrace-diffusion 
mechanism \cite{PT,Ted,EDW}. For early times, such that 
$t\ll L^z$, the interface width $\langle w_{L}^{2}(t)\rangle $ increases 
with time as 
\begin{equation} 
\langle w_{L}^{2}(t)\rangle \sim t^{2\beta}\;, 
\end{equation} 
where $\beta=(z-1)/2z$ \cite{Barabasi,Krug}. 
In the $t \rightarrow \infty$ limit, where $t\gg L^z$, the interface width 
saturates for a finite system, but diverges with $L$ according to  
$\langle w_{L}^{2}(\infty)\rangle \sim L^{2\alpha}$ 
where $\alpha=(z-1)/2$ is the roughness exponent \cite{Barabasi,Krug}. 
 
For $z=4$ (curvature driven interface) we saw that the slope fluctuation  
behaves as $C^{\phi}_{L}(0)=\langle\phi_{i}^2\rangle \sim L$. For higher $z$  
for the slope-slope correlation function one can deduce 
\begin{equation} 
C^{\phi}_{L}(l) = 
\frac{D}{L}\sum_{n=1}^{L-1} \frac{e^{i \frac{2\pi n}{L} l}} 
{\nu \left[2\left(1-\cos\left(\frac{2\pi n}{L}\right)\right) 
\right]^{\frac{z-2}{2}}}  
\;. 
\end{equation} 
It is divergent in the $L \rightarrow \infty$ limit, as a result of 
infinitely small wave-vectors $\sim 1/L$, and we can see that  
\begin{equation} 
C^{\phi}_{L}(0)\sim L^{z-3} \;.  
\label{C_scaling} 
\end{equation} 
It is also useful to define the slope difference correlation function 
\begin{equation} 
G^{\phi}_{L}(l) \equiv  \langle(\phi_{i+l} -  \phi_{i})^2 \rangle 
\end{equation} 
for which one can write 
\begin{equation} 
G^{\phi}_{L}(l) = 
\frac{D}{L}\sum_{n=1}^{L-1} \frac{2\left(1-\cos\left(\frac{2\pi n}{L}l\right)\right)} 
{\nu \left[2\left(1-\cos\left(\frac{2\pi n}{L}\right)\right) 
\right]^{\frac{z-2}{2}}} 
\;. 
\end{equation} 
For the small wave-vector behavior we can again deduce that for $z>5$ 
\begin{equation} 
G^{\phi}_{L}(l) \sim L^{z-5} \, l^2 \;. 
\label{anomalous} 
\end{equation} 
One may refer to this form as ``anomalous'' scaling \cite{Krug} for the slope  
difference correlation function  in the following sense.  
For $z<5$ the scaling form for $G^{\phi}_{L}(l)$ follows that of  
$C^{\phi}_{L}(0)$ [eq. (\ref{C_scaling})], i.e.,  
$G^{\phi}_{L}(l) \sim l^{z-3}$. For $z>5$ [eq. (\ref{anomalous})] it obviously  
features a different $l$ dependence and an additional power of $L$, and it 
diverges in the $L \rightarrow \infty$ limit. 
 
Having these scaling functions for large $L$, we can easily obtain the 
scaling behavior for the average density of local minima. Exploiting the 
identity 
\begin{equation} 
C^{\phi}_{L}(l) = C^{\phi}_{L}(0) - \frac{1}{2}G^{\phi}_{L}(l) 
\end{equation}  
we use the general form for the local minimum density: 
\begin{equation} 
\langle u\rangle = \frac{1}{2\pi}\arccos\left(  
\frac{C^{\phi}_{L}(1)}{C^{\phi}_{L}(0)} \right) = 
\frac{1}{2\pi}\arccos\left( 1-\frac{1}{2} 
\frac{G^{\phi}_{L}(1)}{C^{\phi}_{L}(0)} \right) \simeq 
\frac{1}{2\pi}\arccos\left( 1-\frac{\rm const.}{L^2} \right) \sim 
\frac{1}{L} 
\end{equation} 
Note that this is the scaling behavior for {\em all} $z>5$. 
It simply shows the trivial lower bound for $\langle u\rangle$: since there is 
always at least one minima (and one maxima) among the $L$ sites, it can 
never be smaller than $1/L$. 
 
\subsection{The average curvature at local minima} 

The next natural question to ask is how the average curvature, $K$ at the  
minimum points scales with the system size for the general system  
described by eq. (\ref{kinetic}). This can be evaluated as the  
conditional average of the local curvature at the local minima: 
\begin{equation} 
\langle K\rangle_{\min} = \langle (\phi_{i}-\phi_{i-1}) \rangle_{\min} = 
\frac{\langle (\phi_{i}-\phi_{i-1})\Theta(-\phi_{i-1})  
\Theta(\phi_{i})\rangle }{\langle\Theta(-\phi_{i-1}) 
\Theta(\phi_{i})\rangle} = 
\frac{\langle (\phi_{2}-\phi_{1})\Theta(-\phi_{1}) \Theta(\phi_{2})\rangle 
} 
{\langle u \rangle} 
\label{curvature} 
\end{equation} 
where translational invariance is exploited again. 
The numerator in (\ref{curvature}) can be obtained after performing the 
same basis transformation (Appendix A) that was essential to find  
$\langle u \rangle$. Then after elementary integrations we find 
\begin{equation} 
\langle K\rangle_{\min} = \frac{1}{\langle u \rangle} 
\frac{1}{\sqrt{2\pi}}  
\frac{C^{\phi}_{L}(0)-C^{\phi}_{L}(1)}{\sqrt{C^{\phi}_{L}(0)}} = 
\frac{\sqrt{2\pi}}{\sqrt{C^{\phi}_{L}(0)}}\; 
\frac{C^{\phi}_{L}(0)-C^{\phi}_{L}(1)}{  
\arccos(C^{\phi}_{L}(1)/C^{\phi}_{L}(0))} 
\end{equation} 
Using the explicit results for the slope correlation function for $z=2$ 
and  
$z=4$, and the scaling forms for it for higher $z$ given in the previous  
subsections, one can easily deduce the following. 
For $z<5$ the average curvature at the local minimum points on a lattice 
tends 
to a {\em constant} in the thermodynamic limit. For $z=2$ 
\begin{equation} 
\langle K\rangle_{\min} \simeq  
\frac{2\sqrt{2}}{\sqrt{\pi}}\sqrt{ \frac{D}{\nu}}  
+ {\cal O}\left(\frac{1}{L}\right)\;, 
\end{equation} 
and for $z=4$ 
\begin{equation} 
\langle K\rangle_{\min} \simeq \sqrt{\pi}\sqrt{ \frac{D}{2\nu}} 
+ {\cal O}\left(\frac{1}{L}\right)\;. \label{ke4}
\end{equation} 
The behavior of this quantity drastically changes for $z>5$, where it  
{\em diverges} with the system size as: 
\begin{equation} 
\langle K\rangle_{\min} \sim L^{\frac{z-5}{2}} 
\end{equation}. 
 
\section{Other lattice models and an application to parallel computing} 

\subsection{The single-step model} 
 
In the single-step model the height differences (i.e., the local slopes) are  
restricted to $\pm 1$, and the evolution consists of particles of height $2$  
being deposited at the local minima. While the full dynamic behavior of the  
model belongs to the KPZ universality class, in one dimension the steady state 
is governed by the EW Hamiltonian \cite{single_step}. Thus, the roughness  
exponent is $\alpha=1/2$, and we expect the finite-size effects for  
$\langle u\rangle$ to follow eq. (\ref{EW2_min_dens}).    
The advantage of this model is that 
it can be mapped onto a hard-core lattice gas for which the {\em steady-state} 
probability distribution of the configurations is known exactly  
\cite{single_step,Spitzer}. This enables us to 
find arbitrary moments of the local minimum density operator. 
Since $\phi_i=\pm 1$, it can be simly written as 
\begin{equation} 
u = \frac{1}{L} \sum_{i=1}^{L}\frac{1-\phi_{i-1}}{2}\; \frac{1+\phi_i}{2} = 
    \frac{1}{L} \sum_{i=1}^{L} (1-n_{i-1}) n_i\;, 
\label{ss_min_dens_op} 
\end{equation} 
where $n_i=(1+\phi_i)/2$, corresponds to the hard core lattice gas occupation 
number. The constraint $\sum_{i=1}^{L}\phi_i=0$ translates to  
$\sum_{i=1}^{L}n_i=L/2$. 
Note that here $\langle u\rangle=\langle(1-n_{i-1})n_i\rangle$ is proportional  
to the average current. Knowing the exact steady-state probability  
distribution \cite{single_step,Spitzer}, one can easily find that 
\begin{equation} 
\langle n_i \rangle= \frac{1}{2} 
\;,\;\;\; 
\langle n_i n_j \rangle_{i\neq j}  
                                  = \frac{1}{4}\;\frac{L-2}{L-1} 
\end{equation} 
Thus the exact finite-size effects for the local minimum density: 
\begin{equation} 
\langle u\rangle_L =  \frac{1}{4}\;\frac{L}{L-1} 
= \frac{1}{4} + \frac{1}{4L} + {\cal O}(L^{-2})\;, 
\end{equation} 
in qualitative agreement with (\ref{ezaz}).

\subsection{The Massively Parallel Exponential Update model}

One of the most challenging areas in parallel computing \cite{parallel} 
is the efficient implementation of dynamic Monte-Carlo algorithms 
for discrete-event simulations on massively parallel architectures. 
As already mentioned in the Introduction, it has numerous practical 
applications ranging from magnetic systems (the discrete events 
are spin flips) to queueing networks ( the discrete events are job arrivals). 
 A parallel architecture by definition  contains (usually) a large number of  
processors, or processing elements (PE-s). During the simulation 
each processor has to tackle only a fraction of the full computing 
task (e.g., a  specific block of spins), and the algorithm has to ensure 
through synchronization that the underlying dynamics is not altered. 
In a wide range of models the discrete events are Poisson arrivals. Since this 
stochastic process is reproducible (the sum of two Poisson processes is a  
Poisson process again with a new arrival frequency), the Poisson streams 
can be simulated simultaneously on each subsystem carried by each PE. 
As a consequence, the simulated time is {\em local} and {\em random},  
incremented by exponentially distributed random variables on each PE. 
However, the algorithm has to ensure that causality accross the boundaries of  
the neighboring blocks is not violated. This 
requires a comparison between the neighboring simulated times, and waiting, if 
necessary (conservative approach). In the simplest scenario (one site/PE), 
this means that only those PEs will be allowed to attempt the update the state 
of the underlying site and increment their local time, where the local  
simulated time is a {\em local minimum} regarding the full simulated time  
horizon of the system, $\{\tau_i\}$, $i=1,..,L$ (for simplicity we consider a  
chain-like connectivity among the PE-s but connectivities of higher degree can 
be treated as well). One can in fact think of the time horizon as a  
fluctuating surface with height variable $\tau_i$.  
Other examples where the update attempts are independent Poisson arrivals 
include arriving calls 
in the wireless cellular network of a large metropolitan area \cite{GLNW}, or 
the spin flip attempts in an Ising ferromagnet. This  extremely robust parallel 
scheme was introduced by Lubachevsky, \cite{Luba} and it is applicable to a 
wide range of stochastic cellular automata with local  dynamics where the 
discrete events are Poisson arrivals. The local random time increments is, 
in the language of the associated surface, equivalent to depositing random 
amounts of `material' (with an exponential distribution) at the local minima 
of the surface, see Figure 2. This in fact defines a simple surface growth 
model which we shall refer to as `the massively parallel exponential update 
model' (MPEU). 
The main concern about a parallel implementation is its efficiency. Since 
in the next time step only a fraction of PE-s will get updated, i.e., those 
that are in the local minima of the time horizon, while the rest are in idle, 
the efficiency is nothing but the average number of non-idling PE-s divided by 
the total number of PE-s ($L$), i.e., {\em the average number of minima per 
unit length}, or the minimum-point density, $u$.  
The fundamental question of the  so called {\em scalability} arises: will the 
efficiency of the algorithm go to zero as the number of PE-s is increased 
($L\to \infty$) indefinitely, or not? If the efficiency has a non-zero lower 
bound for $L\to \infty$ the algorithm is called {\em scalable}, and 
certainly this is the preferred type of scheme. Can one design in principle 
such efficient algorithms? 

\begin{figure}
\hspace*{3.5cm}\epsfxsize=4 in  
\epsfbox{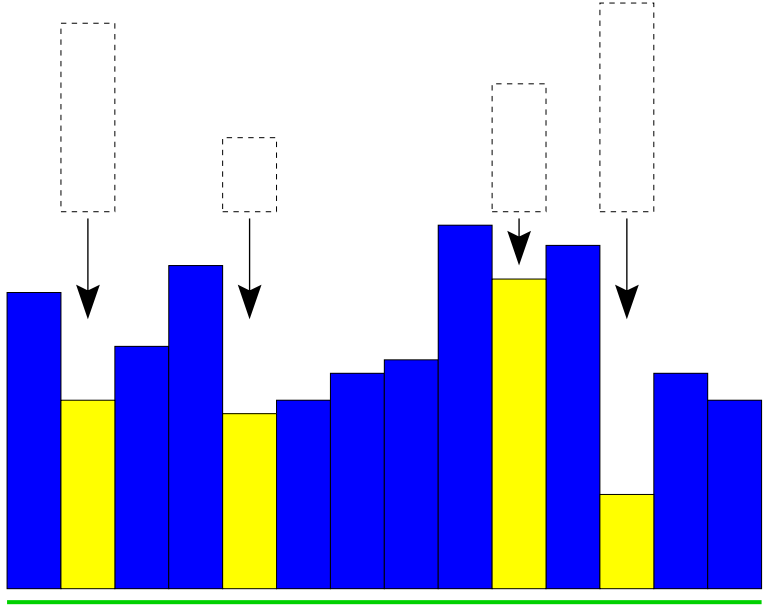} 
\vspace*{0.5cm} 
\caption{The MPEU model. The arrows show the local minima where 
random amounts of material will be deposited in the next time step.} 
\end{figure} 

As mentioned in the Introduction, we know of one
example  that nature provides with an efficient algorithm for a very large
number  of processing elements: the human brain with its $10^{11}$ PE-s is the
largest  parallel computer ever built. Although the intuition suggests that
indeed there  are scalable parallel schemes, it has only been proved recently,
see for  details Ref. \cite{KTNR}, by using the aforementioned analogy with
the  simple MPEU surface growth model. 
While the MPEU model  {\em exactly} mimics the evolution of the simulated  
time-horizon, it can also be considered as a primitive model for ion  
sputtering of surfaces (etching dynamics): to see this, define a new height  
variable via $h_i \equiv -\tau_i$, i.e. flip Figure 2 upside down. This means 
that instead of depositing material we have to take, `etch', and this has to 
be done at the local {\em maxima} of the $\{h_i\}$ surface. In sputtering 
of surfaces by ion bombardement an incoming ion-projectile will most likely 
`break off' a piece from the top of a mound instead from a valley, very 
similar to our `reversed' MPEU model. It was shown that the sputtering process 
is described by the KPZ equation, \cite{bru,Barabasi}. 
This qualitative argument is in complete agrrement with the extensive MC  
simulations and a coarse-grained approximation of Ref. \cite{KTNR} that MPEU, 
similar to the  single-step model, it also belongs to the KPZ dynamic 
universality class; in one dimension the macroscopic landscape is governed by 
the EW  Hamiltonian.   
 
The slope varaibles $\phi_i$ for MPEU are not independent in the  
$L\rightarrow\infty$ limit, but short-ranged. This already guarantees that 
the steady-state behavior is governed by the EW Hamiltonian, and 
the density of local minima does not vanish in the thermodynamic limit.  
Our results confirm that the finite-size effects for $\langle u\rangle$ 
follow eq. (\ref{EW2_min_dens}): 
\begin{equation} 
\langle u\rangle_L \simeq \langle u\rangle_{\infty} + \frac{\rm const.}{L} 
\end{equation}  
with $\langle u\rangle_{\infty}=0.24641(7)$, see Fig. 3. 

\begin{figure}
\hspace*{3.5cm}\epsfxsize=4 in  
\epsfbox{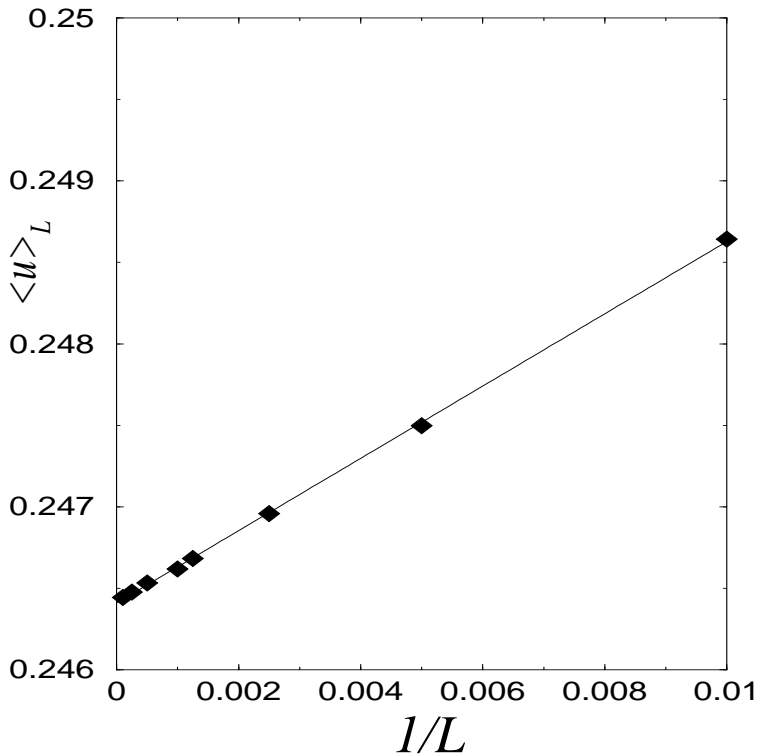} 
\vspace*{0.5cm} 
\caption{Density of minima vs. $1/L$ for the MPEU model.} 
 \end{figure}

We conclude that the basic algorithm 
(one site per PE) is scalable for one-dimensional arrays. The same  
correspondence can be applied to model the performance of the algorithm for 
higher-dimensional logical PE topologies. While this will involve the typical 
difficulties of surface-growth modeling, such as an absence of exact results  
and very long simulation times, it establishes potentially fruitful  
connections between two traditionally separate research areas.   
 
\subsection{The larger curvature model} 

In this subsection we briefly present a curvature driven SOS surface
deposition model known in the literature as the larger curvature model,
and show a numerical analysis of the density of minima on this model.
 This model was originally introduced by Kim and Das Sarma \cite{KSDS}
and Krug \cite{Krug2} independently, as an atomistic deposition
model which fully  conforms to the behaviour of the
continuum fourth order linear Mullins equation ($\nu=0$, $\kappa > 0$ in
Eq. \ref{h_evolution}). Note that the discrete analysis we presented in
Section II is based on the discretization of the continuum equation
using the simplest forward Euler differencing scheme. The larger curvature
model, however, is a {\em growth} model where the freshly deposited particles
diffuse  on the surface according to the rules of the model until they are 
embedded. Since in all the quantities studied so far, the correspondence
(on the level of scaling) between the larger curvature model and the Langevin
equation is very good, we would expect that the dynamic scaling properties of
the  density of minima for both the model and the equation to be identical. 

 The large curvature model has rather simple rules: a freshly deposited atom
(let us say at site $i$) will be incorporated at the nearest neighbor site
which has largest curvature (i.e., $K_i=h_{i+1}+h_{i-1}-2h_i$ is maximum).
If there are more neighbors with the same maximum curvature, then 
one is chosen randomly. If the original site ($i$) 
is among those with maximum curvature, then the atom is incorporated at $i$.

Figure 4 shows the scaling of the density of minima $<u>_L$ in the steady
state, vs. $1/\sqrt{L}$. According to Eq. (\ref{u4d}), for the fourth order
equation on the lattice, the behavior of the density of minima in the steady
state scales with system size as $1/\sqrt{L}$. And indeed, Figure 4 shows the
same behaviour for the larger curvature model, as expected. Note that this
behavior sets in at rather small system sizes already, at about $L=100$,
meaning that the finite system size effects are rather small for the larger
curvature model. This is a very fortunate property since increasing the system
size means decreasing the density of minima, therefore relative statistical
errors will increase. 

\begin{figure}
\hspace*{3.5cm}\epsfxsize=4 in  
\epsfbox{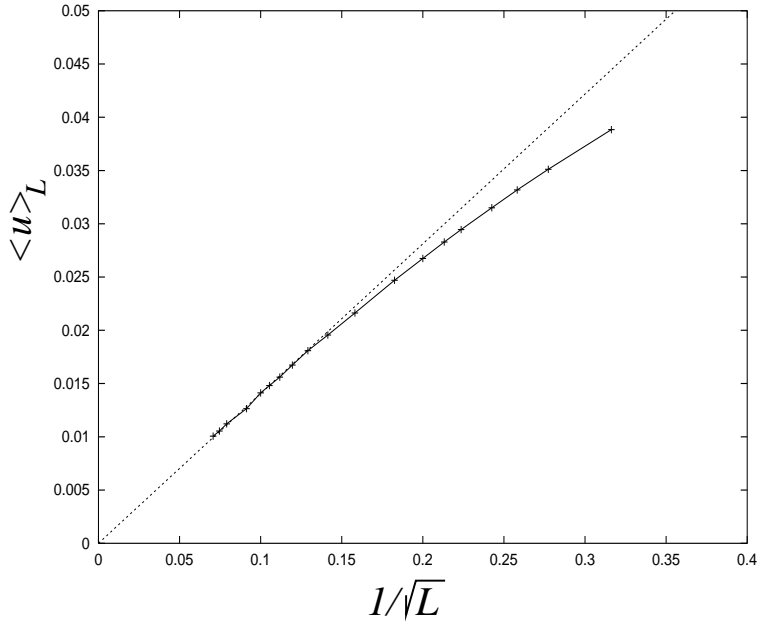} 
\vspace*{0.5cm} 
\caption{Density of minima in the steady state for the larger curvature
model.}  \end{figure} 

This can only be improved by better statistics, i.e.,
with averages over larger number of runs. This becomes however quickly
a daunting task, since the cross-over time toward the steady state scales
with system size as $L^4$.  As we shall see in Section V.A, a matematically
rigorous approach to the continuum equation yields the same $1/\sqrt{L}$
behaviour. Since the density of minima does decay to zero, an algorithm
corresponding to the larger curvature model (or the Mullins equation)
would {\em not} be asymptotically scalable.

Finally, we would like to make a brief note 
about the observed morphologies in the steady state for the Mullins equation
, or the related models.
It has been shown previously \cite{ALF} that in the steady state the
morphology tipically shows a single large mound (or macroscopic groove).
At first sight this may appear as a surprise, since we have shown that the
number of minima (or maxima) diverges as $\sqrt{L}$ (the density vanishes as
$1/\sqrt{L}$. There is however no contradiction, because that refers to a
a mound that expands throughout the system, i.e. it is a long wavelenght
structure, whereas the number of minima measures {\em all} the minima, and
thus it is a short wavelenght characteristic. In the steady state we indeed
have a single large, macroscopic groove, however, there are numerous small
dips and humps generated by the constant coupling to the noise.

\section{Extremal-point densities on the continuum} 
 
 Let us consider a continuous and at least two times differentiable function 
$f:[0,L] \to \R$. We are interested in counting the total number of extrema 
of $f$ in the $[0,L] $ interval. The 
topology of continuous curves in one dimension allows for three possibilities 
on the nature of a point $x_i$ for which $f'|_{x_i} = 0 $. 
Namely, $x_i$ is a local minimum  if  $f''|_{x_i}>0 $,
a local maximum if  $f''|_{x_i}<0$ and it is degenerate if
$f''|_{x_i}=0$. We call the point
$x_i$ a degenerate flat of order $k$,  if $f^{(j)}|_{x_i} = 0$ for
$j=1,2,..,k$ and $f^{(k+1)}|_{x_i} \neq 0$,   $k\geq 2$, assuming that the
higher order derivatives   $f^{(j)}$ implied exist.  The counter-like quantity 
\begin{equation} 
c(L,[f]) \equiv  \frac{1}{L} \int\limits_0^L dx \; |f''|\; \delta(f')\label{counter}    
\end{equation}  
where $\delta$ is the Dirac-delta,  
gives the number of extremum points per unit 
length in the interval $[0,L]$, which in the limit of $L \to 0$ is the 
extremum point density of $f$ in the origin. 
For our purposes $L$ will always be a finite number, however, for the sake
of briefness we shall refer to $c$ simply as the density of extrema.
Note that counting the extrema
of a function  $f$ is equivalent to counting the zeros of its derivate $f'$.
The divergence  of $c$ for finite $L$ implies either the existence of
completely flat regions (infinitely degenerate), 
or an ``infinitely wrinkled'' region, such as for the truncated
Weierstrass function  shown in Fig. 2 ( in this latter case the divergence is
understood by taking  the limit $M \to \infty$). As already explained in the
Introduction this infinitely wrinkled region does not necessarily imply
that the curve is fractal, but if the curve is  fractal, then regions of
infinite wrinkledness must exist. The divergence or non-divergence of
$c$ can be used as an indicator of the existence of such regions (completely
flat or infinitely wrinkled).

One can make the following precise statement  related to the counter $c$: 
if $x_i$ is an extremum point of $f$ of at most  finite degeneracy $k$,  and if
there exist a small enough $\epsilon$,  such that  $f$ is 
analytic in the neighborhood $[x_i-\epsilon, x_i+\epsilon]$,  and there 
are no other extrema in this neighborhood, then 
\begin{equation} 
I(x_i) \equiv \int\limits_{x_i-\epsilon}^{x_i+\epsilon}dx\; |f''|\; 
\delta(f') = 1\;,\;\;\;0 < \epsilon \ll 1 \label{stat1} 
\end{equation} 
In the following we give a proof to this statement.  
 
Using Taylor-series expansions around $x_i$, one writes: 
\begin{eqnarray} 
&&f'(x)=\frac{a_k}{k!} (x-x_i)^k+ 
\frac{a_{k+1}}{(k+1)!} (x-x_i)^{k+1}+... \label{exp1}\\ 
&& f''(x)=\frac{a_k}{(k-1)!} (x-x_i)^{k-1}+ 
\frac{a_{k+1}}{k!} (x-x_i)^{k}+...\label{exp2} 
\end{eqnarray} 
where we introduced the shorthand notation $a_j \equiv f^{(j+1)}|_{x_i}$. 
For the non-degenerate case of $k=1$,  (\ref{stat1})  
follows from a classical property of the 
delta function, namely:  
\begin{equation} 
\delta(g(x)) =  
\sum_i |g'(x_i)|^{-1} \; \delta(x-x_i)\; 
, \;\;\;\mbox{ $x_i$ are {\em simple}  
zeros of $g$}. \label{dprop} 
\end{equation} 
Let us now assume that $x_i$ is degenerate of order $k$ ($k \geq 2$). 
Using the expansions (\ref{exp1}),  (\ref{exp2}),  the variable change  
$u=x-x_i$, and the 
well-known  property $\delta(ax)=|a|^{-1} \delta(x)$, we obtain: 
\begin{equation} 
I(x_i) = k \int\limits_{-\epsilon}^{\epsilon} du\; |u|^{k-1}\; 
\left|1+\sum_{j=1}^{\infty} \frac{(k-1)!}{(k-1+j)!}\; 
\frac{a_{k+j}}{a_k}\;u^j \right|\;\delta\left( |u|^{k} 
\left[  1+\sum_{j=1}^{\infty} \frac{k!}{(k+j)!}\; 
\frac{a_{k+j}}{a_k}\;u^j\right]\right) \label{31} 
\end{equation} 
Next we split the integral (\ref{31}) in two:  
$\int_{-\epsilon}^{\epsilon} ... = \int_{-\epsilon}^{0} ... 
+\int_{0}^{\epsilon}$, make the variable change $u \to -u$ 
in the first one, and then $z=u^k$ in both integrals. The final 
expression can then be written in the form: 
\begin{equation} 
I(x_i) = \int\limits_{-\epsilon^k}^{\epsilon^k} dz\; 
|A(z)|\;\delta(zB(z)) \label{33}\;, 
 \end{equation} 
where 
\begin{equation} 
A(z)=1+\sum_{j=1}^{\infty} \frac{(k-1)!}{(k-1+j)!}\; 
\frac{a_{k+j}}{a_k}\;z^j |z|^{\frac{j}{k}-j}\;,\;\;\;\mbox{and}\;\;\; 
B(z)=1+\sum_{j=1}^{\infty} \frac{k!}{(k+j)!}\; 
\frac{a_{k+j}}{a_k}\;z^j |z|^{\frac{j}{k}-j} \label{34} 
\end{equation} 
We have $A(0)=B(0)=1$, and 
\begin{equation} 
[zB(z)]' = 1 +\sum_{j=1}^{\infty} \frac{k!}{(k+j)!}\; 
\frac{a_{k+j}}{a_k}\;\left(\frac{j}{k}+1\right) z^j |z|^{\frac{j}{k}-j} 
\;,\;\;\;\rightarrow\;\;\;\left.[zB(z)]'\right|_{z=0} = 1\;. 
\end{equation} 
(Take the derivatives separately to the right and to the left of $z=0$). 
Thus, since $z=0$ is a {\em simple} zero of $zB(z)$, property  
(\ref{dprop}) can be applied for sufficiently small $\epsilon$: 
\begin{equation} 
I(x_i) = |A(0)|=1 
\end{equation} 
proving our assertion. 
Note that because of 
(\ref{stat1}),  $c$ counts {\em all} the non-degenerate and the finitely 
degenerate points as well,  giving the equal weight of unity to each. Can we 
count the non-degenerate extrema separately? The answer is affirmative, if 
one considers instead of (\ref{counter}) the following quantity: 
\begin{equation} 
c_q(L,[f]) \equiv \frac{1}{L} \int\limits_0^L dx \; |f''|^{q+1}\;\delta (f') 
\;,\;\;\;q > 0\label{counterq}  
\end{equation} 
Performing the same steps as above we obtain for a degenerate point: 
\begin{equation} 
I_q(x_i) \equiv \int\limits_{x_i-\epsilon}^{x_i+\epsilon}dx\; |f''|^{q+1}\; 
\delta (f') = \left[ \frac{|a_k|}{(k-1)!}\right]^q  
\int\limits_{-\epsilon^k}^{\epsilon^k} dz\; |z|^{q\left(1-\frac{1}{k}\right)} 
\;|A(z)|^{q+1}\;\delta(zB(z)) \label{37}\;.\label{stat2} 
\end{equation} 
Since $k \geq 2$, $q\left(1-\frac{1}{k}\right) \geq \frac{1}{2}q > 0$, 
i.e.,  
\begin{equation} 
I_q(x_i) =0\;,\;\;\;\mbox{for $x_i$ degenerate}. 
\end{equation} 
This means, that $q > 0$ eliminates the degenerate points from the count. 
To non-degenerate points ($k=1$)  (\ref{counterq}) gives the weight of  
\begin{equation} 
I_q(x_i) =|a_1|^q = \Big|f''|_{x_i}\Big|^q\;,\;\;\; 
\mbox{for $x_i$ 
non-degenerate}.  
\end{equation} 
In other words, 
\begin{equation} 
c_q(L,[f])  
=\frac{1}{L} \sum_i \left| K(x_i) \right|^q\;,\;\;\; 
\mbox{$q>0$, $x_i$ non-degenerate 
extrema of $f$} \label{40} 
\end{equation} 
where $K(x) = f''$ is the {\em curvature} of $f$ at $x$.  
The limit $q\to 0^+$ in (\ref{40})  gives the 
extremum point density  $\overline{c}(L,[f])$  
of $f$ of non-degenerate extrema: 
\begin{equation} 
\overline{c}(L,[f]) = \lim_{q \to 0^+} c_q(L,[f]) = \lim_{q \to 0^+}  
\frac{1}{L} \int\limits_0^L dx \; |f''|^{q+1}\;\delta (f') \label{42} 
\end{equation} 
It is important 
to note, that taking the $q \to 0^+$ limit in (\ref{40}) {\em is not  
equivalent} to taking $q=0$ in (\ref{counterq}), i.e., the 
limit and the integral on the rhs of (\ref{42}) are not 
interchangeable! The difference is 
the set of degenerate points!  
 
Until now, we did not make any distiction between maxima and minima. 
In a natural way, we expect that the quantity: 
\begin{equation} 
u(L,[f]) \equiv \frac{1}{L} \int\limits_0^L dx \; f''\; 
\delta (f')\; \theta(f'') \label{43} 
\end{equation} 
where $\theta(x)$ is the Heaviside step-function, will give the 
density of minima (due to the step function,  
here we can drop the absolute values).  
However, performing a similar derivation as above, one concludes 
that (\ref{43}) is a little bit ill-defined, in the sense that the weight 
given to degenerate points depends on the definition of the step-function in 
the origin (however, $u(L,[f])$ is bounded). Introducing a $q-regulator$ as above, 
the weight of degenerate points is pulled down to zero: 
\begin{equation} 
u_q(L,[f]) \equiv \frac{1}{L} \int\limits_0^L dx \; [f'']^{q+1}\; 
\delta (f')\; \theta(f'') \;,\;\;\;q>0.\label{44} 
\end{equation} 
and 
\begin{equation} 
u_q(L,[f]) =\frac{1}{L} \sum_i \left[ K(x_i) \right]^q\;,\;\;\; 
\mbox{$q>0$, $x_i$ non-degenerate 
minima of $f$} \label{45} 
\end{equation} 
Note that in the equation above the absolute values are not needed, since  
we are summing over the curvatures of all local {\em minima}. 
The density $\overline{u}(L,[f]) $  
of non-degenerate minima of $f$ is obtained 
after taking the limit $q\to 0^+$: 
\begin{equation} 
\overline{u}(L,[f]) = \lim_{q\to 0^+} u_q(L,[f])  \label{47} 
\end{equation} 
and the limit is not interchangeable with the integral in (\ref{44}). 
To obtain densities for maxima, one only has to replace 
the argument $f''$  of the Heaviside function with $-f''$. 
 
\subsection{Stochastic extremal-point densities} 
 
We are interested to explore the previously introduced quantities 
for a stochastic function, subject to time evolution, $h(x,t)$.  
This function may be for example the solution to a Langevin equation. 
We define the two basic quantities in the same way as before, 
except that now one performs a stochastic average over the noise, 
as well: 
\begin{eqnarray} 
C_q(L,t)= \left\langle   
\frac{1}{L} \int\limits_0^L dx \;  
\left|\frac{\partial^2 h}{\partial x^2}\right|^{q+1} 
\delta \left( \frac{\partial h}{\partial x}\right)\right\rangle  
\;,\;\;\;\mbox{and}\;\;\; 
U_q(L,t)= \left\langle   
\frac{1}{L} \int\limits_0^L dx \;  
\left[\frac{\partial^2 h}{\partial x^2}\right]^{q+1} 
\delta \left( \frac{\partial h}{\partial x}\right)\; 
\theta \left( \frac{\partial^2 h}{\partial x^2}\right) 
\right\rangle  
\end{eqnarray} 
For systems preserving translational invariance, the stochastic 
average of the integrand becomes $x$-independent, and the integrals 
can be dropped: 
\begin{eqnarray} 
& &C_q(L,t)= \left\langle   
\left|\frac{\partial^2 h}{\partial x^2}\right|^{q+1} 
\delta \left( \frac{\partial h}{\partial x}\right)\right\rangle  \label{Cq}\\ 
& &U_q(L,t)= \left\langle   
\left[\frac{\partial^2 h}{\partial x^2}\right]^{q+1} 
\delta \left( \frac{\partial h}{\partial x}\right)\; 
\theta \left( \frac{\partial^2 h}{\partial x^2}\right) 
\right\rangle \label{Uq} 
\end{eqnarray} 
According to (\ref{40}) and (\ref{45}), $C_q(L,t)$ and $U_q(L,t)$ can be thought of 
as  time dependent ``partition functions'' for the non-degenerate 
extremal-point densities of the underlying stochastic process, with $q$ 
playing the role of  ``inverse temperature'': 
\begin{eqnarray} 
& &C_q(L,t)= \left\langle \frac{1}{L} \sum_i \left| K(x_i) \right|^q 
\right\rangle\;,\;\;\; \mbox{$q>0$, $x_i$ non-degenerate  
extrema} \\ 
& &U_q(L,t)=\left\langle \frac{1}{L} \sum_i \left[ K(x_i) \right]^q 
\right\rangle\;,\;\;\; \mbox{$q>0$, $x_i$ non-degenerate  
minima}  
\end{eqnarray} 
It is important to mention that in the above equations the average $<..>$ 
and the summation are {\em not} interchangeable: particular 
realizations of $h$ have particular sets of minima.  

Two values for $q$ are of special interest: when $q\to 0^+$ 
and $q=1$. In the first case we obtain the stochastic average of the 
density of non-degenerate  extrema and minima: 
\begin{equation} 
\overline{C}(L,t) = \lim_{q\to 0^+} C_q(L,t)\;,\;\;\; 
\mbox{and} \;\;\;\overline{U}(L,t) = \lim_{q\to 0^+} U_q(L,t)\;, \label{bar} 
\end{equation} 
and in the second case we obtain the stochastic average of the mean 
curvature at  extrema and minima: 
\begin{equation} 
\overline{K}_{ext}(L,t) = \frac{C_1(L,t)}{\overline{C}(L,t) }\;,\;\;\; 
\mbox{and} \;\;\;\overline{K}_{min}(L,t) =  
\frac{U_1(L,t)}{\overline{U}(L,t) }\;, \label{curv} 
\end{equation} 
(we need to normalize with the number of extrema/minima per unit 
length to get the curvature per extremum/minimum). 
 
In the following we explore the quantities (\ref{Cq})-(\ref{curv}) 
for a large class of linear Langevin equations. 
To simplify the calculations, we will assume that  
$q$ is a positive integer. Then we will attempt analytic 
continuation on the final result as a function of $q$. In the  
calculations we will make extensive use of the standard integral 
representations of the delta and step functions: 
\begin{eqnarray} 
&&\delta(y) = \int\limits_{-\infty}^{\infty} \frac{dz}{2\pi}\; 
e^{izy} = \sum_{n=0}^{\infty} 
 \int\limits_{-\infty}^{\infty}  
 \frac{dz}{2\pi}\; 
\frac{(iz)^n}{n!}\;y^n\;,  \label{51a} \\ 
&&\theta(y)=\lim_{\epsilon \to 0^+}  
\int\limits_{-\infty}^{\infty} \frac{dz}{2\pi}\; 
\frac{e^{izy} }{\epsilon+iz} = \lim_{\epsilon \to 0^+}  
\sum_{n=0}^{\infty}   \int\limits_{-\infty}^{\infty} \frac{dz}{2\pi}\; 
\frac{1}{\epsilon+iz}\;\frac{(iz)^n}{n!}\;y^n \label{51b} 
\end{eqnarray} 
If $q$ is a positive integer, we may drop the absolute value signs 
in (\ref{Uq}). In (\ref{Cq}) we can only do that for odd $q$. 
The absolute values make the calculation of stochastic averages 
very difficult. We can get around this problem by employing the 
following identity: 
\begin{equation} 
|y|^n = y^n \left\{ (-1)^n  + \theta(y) \left[  
1-(-1)^n 
\right] \right\} \label{52} 
\end{equation} 
This brings (\ref{Cq}) to 
\begin{equation} 
C_q(L,t)=\left[ 1-(-1)^q \right] U_q(L,t)+(-1)^{q+1}B_q(L,t) \label{relation} 
\end{equation} 
where 
\begin{equation} 
B_q(L,t)=\left\langle   
\frac{1}{L} \int\limits_0^L dx \;  
\left(\frac{\partial^2 h}{\partial x^2}\right)^{q+1} 
\delta \left( \frac{\partial h}{\partial x}\right)\right\rangle  
=\left\langle   
\left(\frac{\partial^2 h}{\partial x^2}\right)^{q+1} 
\delta \left( \frac{\partial h}{\partial x}\right)\right\rangle  
\end{equation} 
Obviously for $q$ odd integer, $B_q=C_q$. For $q$ even, 
$B_q$ is an interesting quantity by itself. In this case the 
weight of an extremum  $x_i$ is $sgn (K(x_i)) |K(x_i)|^q$. 
If the analytic continuation can be performed, then the 
$q \to 0^+$ limit will tell us if there are more 
non-degenerate maxima than minima (or otherwise) in average. 
Using the integral representations (\ref{51a}) and (\ref{51b}): 
\begin{eqnarray} 
&&B_q(L,t)=\sum_{n=0}^{\infty} 
 \int\limits_{-\infty}^{\infty}  
 \frac{dz}{2\pi}\;\frac{(iz)^n}{n!}\; 
\left\langle   
\left(\frac{\partial^2 h}{\partial x^2}\right)^{q+1} 
\left( \frac{\partial h}{\partial x}\right)^n 
\right\rangle \label{55a}\\ 
&&U_q(L,t)=\lim_{\epsilon \to 0^+} 
\sum_{n_1=0}^{\infty}\sum_{n_2=0}^{\infty} 
 \int\limits_{-\infty}^{\infty}  
 \frac{dz_1}{2\pi}\;\frac{(iz_1)^{n_1}}{n_1!} 
\int\limits_{-\infty}^{\infty}  
 \frac{dz_2}{2\pi}\;\frac{(iz_2)^{n_2}}{n_2!}\; 
\frac{1}{\epsilon+iz_2}\;
\left\langle   
\left(\frac{\partial^2 h}{\partial x^2}\right)^{n_2+q+1} 
\left( \frac{\partial h}{\partial x}\right)^{n_1} 
\right\rangle \label{55b} 
\end{eqnarray} 
 
\section{Extremal-point densities of linear stochastic evolution 
equations} 
 
 Next we calculate the densities (\ref{55a}), (\ref{55b}) for the 
following type of linear stochastic equations:  
\begin{equation} 
\frac{\partial h}{\partial t} = - \nu \left( - \nabla^2 \right)^{z/2} h + 
\eta(x,t)\;,\;\;\;\;\nu, D, z > 0\;,\;\;\;x\in[0,L] \label{egyenlet} 
\end{equation} 
with initial condition 
$h(x,0)=0$, for all $x \in [0,L]$. 
$\eta$ is a white noise term drawn from a Gaussian distribution with 
zero mean $\langle \eta(x,t) \rangle = 0$, and covariance : 
\begin{equation} 
\langle \eta(x,t)\eta(x',t') \rangle = 2D \delta(x-x')\;\delta(t-t')\;, 
\end{equation} 
We also performed our calculations with other noise types, such as volume 
conserving and long-range correlated, however the details are 
to lengthy to be included in the present paper, it will be the subject
of a future publication. As
boundary condition we choose periodic boundaries:  \begin{eqnarray} 
h(x+nL,t) = h(x,t) \;,\;\;\; 
 \eta(x+nL,t) = \eta(x,t)\;,\;\;\;\mbox{for all $n\in \Z$} 
\end{eqnarray} 
 
The general solution to (\ref{egyenlet}) is obtained simply with the 
help of Fourier series \cite{Krug}.  The Fourier series and its 
coefficients  for  a function $f$ defined on $[0,L]$ is 
\begin{eqnarray} 
f(x)= \sum_{k} \tilde{f}(k)\;e^{ikx}\;,\;\;\;  
\tilde{f}(k)=\frac{1}{L} \int\limits_{-L}^{L} dx\; f(x)\;e^{-ikx} 
\label{Fseries}  
\end{eqnarray} 
where $k=\frac{2\pi}{L}n$, $n= ..,-2,-1,0,1,2,..$. 
The Fourier coefficients of the general solution to (\ref{egyenlet}) 
are: 
\begin{equation} 
\tilde{h}(k,t) = 
\int\limits_{0}^{t} dt'\;e^{-\nu |k|^z (t-t')}\tilde{\eta}(k,t') \label{9} 
\end{equation} 
The correlations of the noise in momentum space are: 
\begin{equation} 
\left\langle \tilde{\eta}(k,t)\tilde{\eta}(k',t')\right\rangle 
=\frac{2D}{L}\;\delta_{k,-k'}\;\delta(t-t')\;. 
\end{equation}

Due to the Gaussian character of the noise, the 
two-point correlation of the solution (\ref{9}) is also delta-correlated and 
it completely  characterizes the statistical 
properties of the stochastic dynamics (\ref{egyenlet}). 
It is given by: 
\begin{equation} 
\left\langle \tilde{h}(k,t)\tilde{h}(k',t')\right\rangle =  
S(k,t) \delta_{k,-k'} \label{hcor} 
\end{equation} 
where $S(k,t)$ is the structure factor given by:\\ 
\begin{equation} 
S(k,t)= 
\frac{D}{\nu L |k|^z}\left[1- e^{-2\nu |k|^zt}\right]\;. \label{sn} 
\end{equation} 
Equation (\ref{egyenlet}) has been analyzed in great detail by a number 
of authors, see Ref. \cite{Krug} for a review. It was shown  
that there exist un upper critical dimension $d_c = z$ for the noisy  
case of Eq. (\ref{egyenlet}) which separates the rough regime with 
$d< z$ from the non-roughening regime $d>z$. In one dimension, 
the rough regime corresponds to the condition $z > 1$, which we 
shall assume from now on, since this is where the interesting physics lies. 
 
 Next, we evaluate the quantities (\ref{Cq})-(\ref{curv}) via directly 
calculating the expressions in (\ref{55a}) and (\ref{55b}). This 
amounts to computing averages of type:  
\begin{equation} 
Q_{N,M} = \left\langle \left( \frac{\partial^2 h}{\partial x^2} 
\right)^N \left( \frac{\partial h}{\partial x} 
\right)^M \right\rangle \label{avg} 
\end{equation} 
Expressing $h$ with its Fourier series according to (\ref{Fseries}), 
we write: 
\begin{eqnarray} 
&&\left( \frac{\partial h}{\partial x} 
\right)^M  = i^M \sum_{k_1}...\sum_{k_M} 
k_1...k_M\; 
\tilde{h}(k_1,t)...\tilde{h}(k_M,t)\;e^{i(k_1+...+k_M)x} \\ 
&&\left( \frac{\partial^2 h}{\partial x^2} 
\right)^N = (-1)^N \sum_{k_1'}...\sum_{k_N'} 
{k_1'}^2...{k_N'}^2\; 
\tilde{h}(k_1',t)...\tilde{h}(k_N',t)\;e^{i(k_1'+...+k_N')x} 
\end{eqnarray} 
which then is inserted in (\ref{avg}). Thus in Fourier  
space one needs to 
calculate averages  of type $\langle \tilde{h}(k_1,t)...\tilde{h}(k_M,t) 
\tilde{h}(k_1',t)...\tilde{h}(k_N',t) \rangle$. According to  
(\ref{hcor}) $\tilde{h}$ is anti-delta-correlated, therefore 
these averages can be performed in the standard way 
\cite{itzi} which is by taking all the possible pairings of indices and 
employing (\ref{hcor}). In our case there are three types 
of pairings: $\{ k_j,k_l\}$, $\{ k_j,k_l'\}$, and $\{ k_j',k_l'\}$. 
Let us pick a `mixed' pair $\{ k_j,k_l'\}$ containing a primed and 
a non-primed index. The corresponding contribution in the 
$Q_{N,M}$ will be: 
\begin{equation} 
\sum_{k_j}\sum_{k_l'} k_j {k_l'}^2 S(k_l',t)  
e^{i(k_j+k_l')x} \delta_{k_j,-k_l'} \label{cont} 
\end{equation} 
Since the structure factor $S(k,t)$  is an even function in 
$k$, (\ref{cont}) becomes  
$\sum_{k_j} {k_j}^3 S(k_j,t) = 0$, because the summand is an odd 
function of $k_j$ and the summation is symmetric around zero. Thus, 
it is enough to consider non-mixed index-pairs, only. This means, that 
$Q_{N,M}$ decouples into: 
\begin{equation} 
Q_{N,M} = \left\langle \left(  
\frac{\partial^2 h}{\partial x^2} 
\right)^N \right\rangle 
\left\langle  
\left( \frac{\partial h}{\partial x} 
\right)^M \right\rangle \label{avg1} 
\end{equation} 
The averages are calculated easily, and we find: 
\begin{eqnarray} 
\left\langle \left( \frac{\partial h}{\partial x} 
\right)^M \right\rangle = \left\{ 
\begin{array}{l} 
(M-1)!! \;\left[ \phi_2(L,t) \right]^{M/2}\;,\;\;\;\mbox{for $M$ even }\;,\\ 
\\ 
0\;,\;\;\;\mbox{for $M$ odd } 
\end{array} 
\right. \label{phi2} 
\end{eqnarray} 
and 
\begin{eqnarray} 
\left\langle \left( \frac{\partial^2 h}{\partial x^2} 
\right)^N \right\rangle = \left\{ 
\begin{array}{l} 
(N-1)!! \;\left[ \phi_4(L,t) \right]^{N/2}\;,\;\;\;\mbox{for $N$ even }\;,\\ 
\\ 
0\;,\;\;\;\mbox{for $M$ odd } 
\end{array} 
\right. \label{phi4} 
\end{eqnarray} 
where 
\begin{equation} 
\phi_m(L,t) \equiv \sum_{k} |k|^m\;S(k,t) \label{phi} 
\end{equation} 
Employing (\ref{phi2}), and (\ref{phi4}) in (\ref{55a}), it follows that 
if $q$ is an even integer, $q=2s$, $s=1,2,..$:  
\begin{equation} 
B_{2s}(t) = 0\;,\;\;\;s=1,2,... \label{77a} 
\end{equation} 
whereas for $q$ odd integer, $q=2s-1$, $s=1,2,..$: 
\begin{equation} 
B_{2s-1}(t) = C_{2s-1}(t) = 
\frac{2^{s-\frac{1}{2}}}{\pi}\;\Gamma 
\left(s+\frac{1}{2}\right)\; 
\frac{\left[ \phi_4(L,t) \right]^{s}}{ 
\sqrt{\phi_2(L,t)}} 
\;,\;\;\;s=1,2,... \label{77b} 
\end{equation} 
where we used the identity $2^p(2p-1)!!/(2p)!=1/p!$, and performed 
the Gaussian integral. 
 
The calculation of $U_q$ is a bit trickier. The sum over $n_1$ in (\ref{55b}) 
is easy and leads to the Gaussian $e^{-\phi_2(L,t) z_1^2 /2}$. However, 
the sum over $n_2$ is more involved. Let us make the temporary 
notation for the sum over $n_2$: 
\begin{equation} 
R_q = \sum_{n_2=0}^{\infty} \frac{(iz_2)^{n_2}}{n_2!}\;(2r-1)!!\; 
\left[ \phi_4(L,t) \right]^{r}\;,\;\;\;n_2+q+1=2r 
\end{equation} 
We have to distinguish two cases according to the parity of $q$:\\ 
1) $q$ is odd, $q=2s-1$, $s=1,2,..$. In this case $R_q$ becomes 
\begin{equation} 
R_{2s-1} = \sum_{r=s}^{\infty} \frac{(iz_2)^{2(r-s)}}{[2(r-s)]!} 
\;\frac{(2r)!}{r!}\; 
\left[ \frac{1}{2}\phi_4(L,t) \right]^{r}=(z_2)^{-2s}(-1)^s 
\left\{ \frac{\partial^{2s}}{\partial x^{2s}} 
\left[ e^{-\phi_4(L,t)z_2^2x^2/2}\right] \right\}_{x=1} 
\end{equation} 
The Hermite polynomials are defined via the Rodrigues formula as: 
\begin{equation} 
H_n(x)  = (-1)^n e^{x^2} \frac{d^n}{dx^n}\left(e^{-x^2}\right) 
\label{Rodrigues} 
\end{equation} 
Using this, we can express $R_{2s-1}$ with the help of Hermite 
polynomials: 
\begin{equation} 
R_{2s-1} =  (-1)^s \left[ \frac{1}{2} \phi_4(L,t) \right]^s \; 
H_{2s}\left( \sqrt{\frac{1}{2} \phi_4(L,t) } z_2 \right)\; 
e^{-\phi_4(L,t)z_2^2x^2/2}   \label{rodd} 
\end{equation} 
\noindent 2) $q$ is even, $q=2s$, $s=1,2,..$. The calculations are analogous 
to the odd case: 
\begin{equation} 
R_{2s} = \sum_{r=s+1}^{\infty} \frac{(iz_2)^{2(r-s)-1}}{[2(r-s)-1]!} 
\;\frac{(2r)!}{r!}\; 
\left[ \frac{1}{2}\phi_4(L,t) \right]^{r}=(iz_2)^{-2s-1} 
\left\{ \frac{\partial^{2s+1}}{\partial x^{2s+1}} 
\left[ e^{-\phi_4(L,t)z_2^2x^2/2}\right] \right\}_{x=1} 
\end{equation} 
or via Hermite polynomials: 
\begin{equation} 
R_{2s} = i (-1)^s \left[ \frac{1}{2} \phi_4(L,t) \right]^{s+\frac{1}{2}} \; 
H_{2s+1}\left( \sqrt{\frac{1}{2} \phi_4(L,t) } z_2 \right)\; 
e^{-\phi_4(L,t)z_2^2x^2/2}   \label{reven} 
\end{equation} 
In order to obtain $U_q$ we have to do the integral over $z_2$ in 
(\ref{55b}). This can be obtained after using the formula: 
\begin{equation} 
\int\limits_{-\infty}^{\infty} dx\;(x\pm ic)^{\nu}\;H_n(x)\;e^{-x^2} =  
2^{n-1-\nu}\sqrt{\pi}\;\frac{\Gamma \left( 
\frac{n-\nu}{2} \right)}{\Gamma(-\nu)}\;e^{\pm \frac{i\pi}{2}(\nu+n)}\;, 
\;\;\;c \to  0^+. 
\end{equation} 
Finally, the densities for the minima read as: 
\begin{eqnarray} 
&&U_{2s}(t)=\frac{2^{s-1}}{\pi}\;\Gamma(s+1)\; 
\frac{\left[ \phi_4(L,t) \right]^{s+\frac{1}{2}}}{ 
\sqrt{\phi_2(L,t)}} \label{77c}\\ 
&&U_{2s-1}(t)=\frac{2^{s-\frac{3}{2}}}{\pi}\;\Gamma\left(s+\frac{1}{2}\right)\; 
\frac{\left[ \phi_4(L,t) \right]^s}{ 
\sqrt{\phi_2(L,t)}} \label{77d} 
\end{eqnarray} 
Formulas (\ref{77a}), (\ref{77b}), (\ref{77c}), and (\ref{77d}) combined with 
(\ref{relation}) 
can be condensed very simply, and we obtain the general result as: 
\begin{eqnarray} 
&&U_q(L,t) = \frac{2^{\frac{q}{2}-1}}{\pi}\; 
\Gamma\left(\frac{q}{2}+1\right)\; 
\frac{\left[ \phi_4(L,t) \right]^{\frac{q+1}{2}}}{ 
\sqrt{\phi_2(L,t)}} \label{78a}\\ 
&&C_q(L,t)=2 U_q(L,t) \label{78b} 
\end{eqnarray} 
Equations (\ref{78a}), (\ref{78b}) with together with (\ref{77d})  
fully solve the problem for the density of non-degenerate 
extrema. Eq. (\ref{78b}) is an expected result in  
one dimension, because Eq. (\ref{egyenlet}) preserves the up-down symmetry. 
The density of non-degenerate minima is: 
\begin{equation} 
\overline{U}(L,t)=\lim_{q\to 0^+} U_q(L,t)=\frac{1}{2 \pi} \; 
\sqrt{\frac{\phi_4(L,t)}{\phi_2(L,t)}} \label{genU} 
\end{equation} 
and the stochastic average of the mean curvature at a minimum point is: 
\begin{equation} 
\overline{K}(L,t)=\frac{U_1(L,t)}{\overline{U}(L,t)}=\sqrt{\frac{\pi}{2}}\; 
\sqrt{\phi_4(L,t)} \label{genK} 
\end{equation} 
i.e., the average curvature at a minimum is proportional to the 
square root of the fourth moment of the structure factor. 
In the following section we exploit the physical information behind 
the above expressions for the  stochastic process (\ref{egyenlet}).  
At some parameter values
a few, or all the quantities above may diverge. In this case 
we introduce a microscopic lattice cut-off $0 < a \ll 1$, and analyze the  
limit $a \to 0^+$ in the final formulas. This in fact corresponds to placing 
the whole problem on a lattice with lattice constant $a$. 
 It has been shown in Ref. \cite{Krug} that for the class  
of equations (\ref{egyenlet}) there are three important length-scales 
that govern the statistical behavior of the interface $h$: the lattice 
constant $a$, the system size $L$, and the {\em dynamical correlation 
length}  $\xi$ defined by: 
\begin{equation} 
\xi(t) \equiv (2\nu t)^{1/z}  \label{xi} 
\end{equation} 
 According to (\ref{phi}) and  (\ref{sn}) the function $\phi_m(L,t)$ becomes:
\bb{ 
\phi_m(L,t)=\frac{2D}{\nu L}\sum_{n=0}^{\infty}  
\left(\frac{2\pi n}{L}\right)^{m-z}\left[1- e^{\textstyle -\left(\xi 
 \frac{2\pi n}{L}\right)^z}\right] 
\;,\;\;\;m=2,4 \label{phin} 
}\e 
The $n=0$ term can be dropped from the sum above, because it is zero 
even for $m < z$ (expand the exponential and then take $n=0$). However, 
the whole sum may diverge depending on $m$ and $z$. In order to handle 
all the cases, including the divergent ones we introduce a microscopic 
lattice cut-off $a$, $0<a \ll 1$, and then analyze the limit $a \to 0^+$ in the 
final expressions. This is in fact equivalent to putting the whole problem on 
a lattice of lattice spacing $a$. Appropiately, (\ref{phin}) becomes: 
\bb{ 
\phi_m(L,t)=\frac{2D}{\nu L}\sum_{n=1}^{\frac{L}{2a}}  
\left(\frac{2\pi n}{L}\right)^{m-z}\left[1- e^{\textstyle -\left(\xi 
 \frac{2\pi n}{L}\right)^z}\right] 
\;,\;\;\;m=2,4 \label{phina} 
}\e 
 
\subsection{Steady-state regime.}  
 
Putting $\xi=\infty$ in (\ref{phina}) $\phi_m$ 
takes a simpler form: 
\bb{ 
\phi_m(L,\infty)=\frac{2D}{\nu L}  
\left(\frac{2\pi}{L}\right)^{m-z} 
\sum_{n=1}^{\frac{L}{2a}}  
n^{m-z} 
\;,\;\;\;m=2,4 \label{phinb} 
}\e 
As $a \to 0^+$, $\phi_m$ becomes proportional to $\zeta(z-m)$. For  
$z-m > 1$ $\phi_m$ is convergent, otherwise it is divergent. In the divergent 
case we quote the following results: 
\bb{ 
\sum_{n=1}^{N} n^s = \ln{N} + {\cal C} + {\cal 
O}(1/N)\;,\;\;\;\mbox{if}\;\;s=-1 \label{euler}  
}\e 
and 
\bb{ 
\sum_{n=1}^{N} n^s =\frac{N^{s+1}}{s+1}\Big[1 + 
{\cal O}(1/N)\Big] \;,\;\;\;\mbox{if}\;\;s > -1 \label{cesaro}  
}\e 
which we will use to derive the leading behaviour of the extremal point 
densities when $L/a \to \infty$. From equations (\ref{phinb}),  
(\ref{78a}), (\ref{genU}) and (\ref{genK}) follows: 
\bb{ 
U_q(L,\infty) = \Gamma\left(\frac{q}{2} +1 \right) 
\left(\frac{2D}{\pi \nu} \right)^{ \frac{q}{2}} \left( 
2\pi \right)^{\frac{q}{2}(5-z)} 
L^{-1-\frac{q}{2}(5-z)} 
\left[ \sum_{n=1}^{L/2a} n^{4-z}\right]^{\frac{q+1}{2}} 
\left[ \sum_{n=1}^{L/2a} n^{2-z}\right]^{-\frac{1}{2}}\;, 
\label{uqst}  
}\e 
\bb{ 
\overline{U}(L,\infty) = \frac{1}{L} \sqrt{  
 \sum_{n=1}^{L/2a} n^{4-z}\left(
\sum_{n=1}^{L/2a} n^{2-z}\right)^{-1} }\;, 
\label{u0st}  
}\e 
and 
\bb{ 
\overline{K}(L,\infty) = \sqrt{\frac{D}{2}  \left( 
2\pi \right)^{5-z} L^{z-5} 
\sum_{n=1}^{L/2a} n^{4-z}}\;.\label{kst}  
}\e 
The convergency (divergency) properties of the sums in 
Eqs. (\ref{uqst}-\ref{kst}) for $a \to 0^+$ generate two critical  
values for $z$, namely $z=3$ and $z=5$. In the three regions 
separated by these values we obtain {\em qualitatively} different 
behaviors for the extremal-point densities.  
 
{\em i)} $z > 5$. All quantities are convergent as $a \to 0^+$. We 
have: \bb{ 
U_q(L,\infty) = \Gamma\left(\frac{q}{2}+1 \right)  
\left(\frac{2D}{\pi \nu} \right)^{\frac{q}{2}} (2\pi)^{\frac{q}{2}(5-z)} 
\frac{[\zeta(z-4)]^{\frac{q+1}{2}}}{[\zeta(z-2)]^{\frac{1}{2}}} 
\;L^{-1+\frac{q}{2}(z-5)}\;,\;\;\;\;z>5 \label{uqzg5} 
}\e 
\bb{ 
\overline{U}(L,\infty) = \frac{1}{L} 
\sqrt{\frac{\zeta(z-4)}{\zeta(z-2)}}\;, \;\;\;\; \label{u0zg5} 
}\e 
\bb{ 
\overline{K}(L,\infty) = (2 \pi)^{\frac{5-z}{2}} 
\sqrt{\frac{D}{2 \nu} 
\zeta(z-4)}\;L^{\frac{z-5}{2}}\;,\;\;\;\; \label{kzg5} 
}\e 
Eq. (\ref{u0zg5}) shows that there are a finite number of minima 
($\sqrt{\zeta(z-4)/\zeta(z-2)}$) in the steady state, independently of the 
system size $L$. ($\overline{U}(L,\infty)$ is the number of minima per unit 
length, and $L\overline{U}(L,\infty)$ is the number of minima on the substrate 
of size $L$). The mean curvature $\overline{K}(L,\infty)$ diverges with 
system size as $L^{(z-5)/2}$. This is consistent with the fact that  
the system size grows as $L$, 
the width grows as $L^{(z-1)/2}$, i.e., faster than $L$, and thus the peaks 
and minima should become sleeker and sharper as $L \to \infty$, expecting 
diverging curvatures in minima and maxima. However, this is not always 
true, since the sleekness of the humps and mounds does not 
necessarily imply large curvatures in minima and maxima if the {\em shape} of 
the humps also changes as $L$ changes, i.e., there is lack of {\em 
self-affinity}. The existence of $z=5$ as a critical value is a non-trivial results 
coming from the presented analysis.  
 
{\em ii)} $z = 5$. According to 
(\ref{euler}), $\phi_4(L,\infty)$ diverges logarithmically as $a \to 0^+$. One 
obtains:  
\bb{ 
U_q(\infty) \simeq \Gamma\left(\frac{q}{2}+1 \right)  
\left(\frac{2D}{\nu \pi} \right)^{\frac{q}{2}} \frac{1}{\sqrt{\zeta(3)}} 
\frac{1}{L} 
\left( \ln{\frac{L}{2a}} + {\cal C}\right)^{\frac{q+1}{2}}\;, 
 \label{uqze5} 
}\e 
\bb{ 
\overline{U}(L,\infty) = \frac{1}{\sqrt{\zeta(3)}}\; 
\frac{1}{L} \sqrt{ \ln{\frac{L}{2a}} + {\cal C}}\; 
\;, \label{u0ze5} 
}\e 
\bb{ 
\overline{K}(L,\infty) =  
\sqrt{\frac{D}{2 \nu} 
\left(  
\ln{\frac{L}{2a}} + {\cal C}\right)}\;. \label{kze5} 
}\e 
 Eq. (\ref{u0ze5}) shows that the although the density of minima 
vanishes, the number of minima is no longer a constant but {\em diverges} 
logarithmically with system size $L$. The mean curvature still diverges, but 
logarithmically, when compared to the power law divergence of (\ref{kzg5}).  
 
For the mean curvature $\overline{K}(\infty)$ in (\ref{kst})  $z=5$ is the only 
critical value, since it only depends on $\phi_4$. For $z < 5$, using Eq.  
(\ref{cesaro}) we arrive to the result that the mean curvature in a minimum 
point approaches to an $L$-independent constant for $L/a \to \infty$ with 
corrections on the order of $a/L$: 
\bb{ 
\overline{K}(L,\infty) \simeq 
\left( \frac{\pi}{a}\right)^{\frac{5-z}{2}} 
\sqrt{\frac{D}{2\nu (5-z)}} \;,  \;\;\;z<5 \label{kzl5}  
}\e 
We arrived to the same conclusion in Section II.D when we studied the
steady state of the discretized version of the continuum equation.
Coincidentally, for $z=4$  the two constant values from (\ref{kzl5})
and (\ref{ke4}) are identical ($a=1$ by definition in (\ref{ke4}).

{\em iii)} $3 < z < 5$.  In this case $\phi_4(L,\infty) \to \infty$ 
and  $\phi_2(L,\infty) < \infty$ as $a \to 0^+$, and: 
\bb{ 
U_q(L,\infty) \simeq \Gamma\left(\frac{q}{2}+1 \right)  
\left(\frac{2D}{\nu \pi} \right)^{\frac{q}{2}}  
\left(\frac{\pi}{a}\right)^{\frac{q}{2}(5-z)} 
\left(\frac{1}{2a}\right)^{\frac{5-z}{2}} 
\frac{L^{-\frac{z-3}{2}}} 
{(5-z)^{\frac{q+1}{2}}\sqrt{\zeta(z-2)}}\;,\label{uq3lzl5} 
}\e 
and  
\bb{ 
\overline{U}(L,\infty) \simeq  \left(\frac{1}{2a}\right)^{\frac{5-z}{2}}\; 
\frac{L^{-\frac{z-3}{2}}}{ \sqrt{ (5-z) \zeta(z-2)}}\; 
\;, \label{u03lzl5} 
}\e 
and the mean curvature is just given by (\ref{kzl5}). 
 
Comparing Eqs. (\ref{uqzg5}), (\ref{uqze5}), and (\ref{uq3lzl5}) we can make 
an interesting observation:  while for $z \geq 5$ the dependence on the system 
size $L$ is coupled to the `inverse temperature' $q$, for $3 < z < 5$ the 
dependence on $L$ {\em decouples} from $q$, i.e., it becomes independent 
of the inverse temperature! Eq. (\ref{u03lzl5}) shows that the density 
of minima vanishes with system size as a power law with an 
exponent $(z-3)/2$ 
but the number of minima of the substrate diverges as a power law with an 
exponent of $(5-z)/2$.  
 
{\em iv)} $z=3$. In this case $\phi_4(L,\infty) \to \infty$ 
and  $\phi_2(L,\infty) \to  \infty$ logarithmically as $a \to 0^+$. 
One obtains: 
\bb{ 
U_q(L,\infty) \simeq  
\frac{1}{2\sqrt{2} a} 
\Gamma\left(\frac{q}{2}+1 \right)  
\left(\frac{\pi D}{\nu a^2} \right)^{\frac{q}{2}}  
\frac{1}{\sqrt{\ln{\frac{L}{2a}}+{\cal C}}} 
\;,\label{uqze3} 
}\e 
and  
\bb{ 
\overline{U}(L,\infty)\simeq  
\frac{1}{2\sqrt{2} a} 
\frac{1}{\sqrt{\ln{\frac{L}{2a}}+{\cal C}}} 
\;,\label{u0ze3} 
}\e 
with a logarithmically vanishing density of minima, and 
the dependence on the 
system size in (\ref{uqze3}) is not coupled to $q$.

{\em v)} $1 < z < 3$. Now both $\phi_4$ and $\phi_2$ 
diverge as $a \to 0^+$. 
Employing (\ref{cesaro}), yields: 
\bb{ 
U_q(L,\infty) \simeq  
\frac{1}{2 a} 
\Gamma\left(\frac{q}{2}+1 \right)  
\left(\frac{2 D}{\pi \nu} \right)^{\frac{q}{2}}  
\left(\frac{\pi}{a}\right)^{\frac{q}{2}(5-z)}  
\frac{\sqrt{3-z}}{(5-z)^{\frac{q+1}{2}}} 
\;,\label{uqzl3} 
}\e 
and  
\bb{ 
\overline{U}(L,\infty)\simeq  
\frac{1}{2 a} \sqrt{ \frac{3-z}{5-z}} 
\;.\label{u0zl3} 
}\e 
Note, that in leading order, both $U_q(L,\infty)$ and the density of minima 
$\overline{U}(L,\infty)$ become system size independent! The system size 
dependence comes in as {\em corrections} on the order of $a/L$ and higher. 
The fact that the efficiency of the massively parallel algorithm presented in 
Section III.B is not vanishing is due precisely to the above phenomenon: 
the fluctuations of the time horizon in the steady state belong to the $z=2$ 
class (Edwards-Wilkinson universality), and according to the results under 
{\em iv)}, the density of minima (or the efficiency of the parallel 
algorithm) converges to a non-zero constant, as $L \to \infty$, ensuring the 
scalability of the algorithm. An algorithm that would map into a $z \geq 3$ 
class would have a vanishing efficiency with increasing the number of 
processing elements. In particular, for $z=2$, one obtains from (\ref{u0zl3}) 
$\overline{U}(L,\infty) \simeq (a 2\sqrt{3})^{-1} = 0.2886.../a$. Note that  
the utilization we obtained is somewhat different from the discrete case 
which was $0.25$. This is due to the fact that this number is non-universal 
and it may show differences depending on the discretization scheme used, 
however it cannot be zero. 
 
Another important conclusion can be drawn from the final results enlisted 
above: at and below $z=5$, all the quantities {\em diverge} when 
$a \to 0^+$, and keep $L$ fixed. This means that the higher the resolution 
the more details we find in the morphology, just as for an
infinitely wrinkled, or a fractal-like  surface. We call this transition
accross $z=5$ a `wrinkle'  transition. As shown in the Introduction,
wrinkledness can assume two phases depending on whether 
the curve is a fractal or not and the transition between these two pases
may be conceived
as  a phase transition. 
However, one may be able to scale the system  size $L$
with $a$ such that the quantities calculated will not diverge in this  limit.
This is possible only in the regime $3< z <5$, when we impose: 
\begin{equation}  
 L^{z-3}a^{(q+1)(5-z)} = const. 
\end{equation} 
This shows that the rescaling cannot be done for all inverse temperatures
$q$ at the same time. 
In particular, for the density of minima and $z=4$,  $La=const$. 
 
\subsection{Scaling regime} 
 
In order to obtain the temporal behavior of the extremal-point densities 
we will use the Poisson summation formula (\ref{ZP}) from Appendix B 
on (\ref{phina}). After simple changes of variables in the integrals 
this leads to:  
\ba{ 
\phi_m(L,t) = \frac{D}{\nu L} \left( \frac{\pi}{a}\right)^{m-z} 
\left[ 1-e^{-\left( \xi \frac{\pi}{a}\right)^{z}}\right] 
+\frac{D}{\pi\nu} \xi^{-(m-z+1)} \int\limits_{0}^{\pi \xi / a} 
dx\; x^{m-z} \left(1-e^{-x^z} \right)+ &&  \nonumber \\ 
\frac{2D}{\pi\nu} \xi^{-(m-z+1)} \sum_{n=1}^{\infty}  
\int\limits_{0}^{\pi\xi / a} dx\; x^{m-z} 
\cos{\left( \frac{L}{\xi}nx\right)} 
 \left(1-e^{-x^z} \right) && \label{phic} 
}\e 
This expression shows, that the scaling properties of the dynamics 
are determined by the dimensionless {\em ratios} $L/\xi$ and $\xi/a$. 
The scaling regime is defined by $a \ll \xi \ll L$.  
 
As we have seen in the previous section, $\phi_m$ is convergent 
for $z > m+1$ but diverges when $z \leq m+1$, as $a \to 0^+$.  
In the convergent case, the lattice spacing $a$ can be taken as zero, and thus 
the first term on the rhs. of (\ref{phic}) vanishes and the time dependence 
of the infinite system-size piece of $\phi_m$ (the first integral term 
in (\ref{phic})) assumes 
the {\em clean} power-law behaviour of $t^{(z-m-1)/z}$ (with a {\em positive} 
exponent). In the divergent case, however, the non-integral term of 
(\ref{phic})  does not vanish, and the time-dependence will not be a clean 
power-law.  
Even the integral terms will present corrections to the 
power-law $t^{-(m+1-z)/z}$ (which has now a {\em negative} exponent), since 
the limits for integration contain $\xi$. The first intergal on the rhs of 
(\ref{phic}) for $z \neq m+1$ can be calculated exactly: 
\bb{ 
\int\limits_{0}^{\pi \xi / a}\!\! 
dx\; x^{m-z} \left(1-e^{-x^z} \right) =  
\frac{1}{m-z+1} \left( \frac{\pi \xi}{a}\right)^{m-z+1}\!\! - 
\frac{1}{z}\Gamma\left( \frac{m-z+1}{z}\right)   
+\frac{1}{z}\Gamma\left( \frac{m-z+1}{z},\left( \frac{\pi 
\xi}{a}\right)^z\right),\;\;z\neq m+1  \label{intt} 
}\e   
where $\Gamma(\alpha,x)$ is the incomplete Gamma function. In our case 
$(\pi \xi / a)^z$ is a large number, and therefore we can use the asymptotic 
representation of $\Gamma(\alpha,x)$ for large $x$, see \cite{RG}, pp. 
951, equation 8.357. According to this, for large $x$,  $\Gamma(\alpha,x) \sim 
x^{\alpha-1}e^{-x}$, i.e., it can become arbitrarily small, with an exponential 
decay. This term can therefore be neglected from (\ref{intt}), compared 
to the other two terms, even in the divergent case.  
Inserting (\ref{intt}) into (\ref{phic}), we will see that also the 
non-integral piece of (\ref{phic}) can be neglected compared to the term 
generated  by the first on the rhs of (\ref{intt}), since in the  
scaling regime $a \ll \xi \ll L$, and thus the ratio $a/L$ can be neglected 
compared to $(m-z+1)^{-1}$. (This is needed only in the divergent 
regime, $z < m+1$.) Thus, one obtains:  
\bb{ 
\phi_m(L,t) \simeq \frac{D}{\pi\nu (m-z+1)} \left( \frac{\pi}{a}\right)^{m-z+1} 
\!\!-\frac{D}{\pi\nu z}  
\Gamma\left( \frac{m-z+1}{z} \right) 
\xi^{z-m-1} \left[ 1- E_m\left( \frac{L}{\xi}, 
\frac{\xi}{a}\right) \right],\;\;z\neq m+1\label{phic1} 
}\e 
where 
\bb{ 
E_m(\lambda,\rho) = \frac{2z}{\Gamma\left( \frac{m-z+1}{z} \right)}  
\sum_{n=1}^{\infty}  
\int\limits_{0}^{\pi\rho} dx\; x^{m-z} 
\cos{( \lambda n x)} 
 \left(1-e^{-x^z} \right),\;\;z\neq m+1 
}\e 
The oscillating terms 
condensed in $E_m$ will give the finite-size corrections, as long as 
$L/\xi \gg 1$.  
 
The $z=m+1$ case (divergent) can also be calculated, however, instead 
of (\ref{intt}) now we have: 
\bb{ 
\int\limits_{0}^{\pi \xi / a}\!\! 
\frac{dx}{x}\;\left(1-e^{-x^z} \right) =  
\ln{\left( \frac{\pi \xi}{a}\right)} +\frac{{\cal C}}{z} -\frac{1}{z} 
\mbox{Ei}\left( - \left( \frac{\pi \xi}{a}\right)^z\right) 
,\;\;z = m+1\;.  \label{intl} 
}\e   
where $\mbox{Ei}(x)$ is the exponential integral function. According to  
the large-$x$ expansion of the exponential integral function, see \cite{RG}, 
pp. 935, equation 8.215, $\mbox{Ei}(-x) \sim -x^{-1}e^{-x}$, it is vanishing 
exponentially fast, thus it can be neglected in the expression of $\phi_m$ 
in the scaling limit: 
\bb{ 
\phi_m(L,t) \simeq  
\frac{D}{\pi \nu} \ln{\left( \frac{\pi \xi}{a}\right)} + 
\frac{D{\cal C}}{\pi\nu z} +  
\frac{D}{\pi \nu} F_m\left( \frac{L}{\xi}, 
\frac{\xi}{a}\right)\;,\;\; z = m+1 \label{phic2} 
}\e 
where 
\bb{ 
F_m(\lambda,\rho) = 2 \sum_{n=1}^{\infty}  
\int\limits_{0}^{\pi\rho} \frac{dx}{x}\;  
\cos{( \lambda n x)} 
 \left(1-e^{-x^z} \right),\;\;z = m+1 
}\e 
 
Thus in the scaling limit, the temporal behavior of $\phi_m$ becomes 
a logarithmic time dependence plus a constant, as long as $L/ \xi \gg 1$. 
 
Observe that for $z < m+1$ the first term 
on the rhs. of (\ref{phic1}), reproduces exactly the diverging term 
(as $a \to 0^+$) of the steady-state expression 
(\ref{phinb}) which can be seen after employing (\ref{euler}) in  (\ref{phinb}). 
This means that for $\xi \to \infty$, $E_m(L/\xi,\xi/a)$ diverges slower than 
$\xi^{m+1-z}$ (this is how the saturation occurs). 
Similarly, for $z = m+1$  
the first term on the rhs of (\ref{phic2})  
(after replacing $\xi$ with $L$ )  
reproduces the diverging term (as $a \to 0^+$) of the steady-state expression 
(\ref{phinb}) which can be seen after employing  
(\ref{cesaro}) in   (\ref{phinb}).  This means, that in the saturation (or 
steady-state) regime the remaining terms from (\ref{phic2}) must behave as 
$const.+{\cal O}(a/L)+\ln{(L/\xi)}$, as $\xi \to \infty$ while keeping $L$ and 
$a$ fixed. 
 
Just as in the case of steady-state one has to distinguish 5 cases  
depending on the values of $z$, with respect to  the critical  
values 3 and 5. For the sake of simplicity of writing,  we 
will omit the arguments of  $E_m(\lambda,\rho)$ and $F_m(\lambda,\rho)$. 
 
{\em i)} $z>5$.  We have: 
\bb{ 
\phi_m(L,t) = \frac{D}{\pi \nu (z-m-1)} \Gamma\left(\frac{m+1}{z}\right)  
\xi^{z-m-1}  (1- E_m)\;,\;\;\;m=2,4 \label{phic3} 
}\e 
From Eqs.  (\ref{78a}), (\ref{genU}) and (\ref{genK}), it follows: 
\bb{ 
U_q(L,t) = \frac{1}{2\pi} \Gamma\left( \frac{q}{2}+1\right) 
\left( \frac{2D}{\pi \nu}\right)^{\frac{q}{2}}  
\left[\frac{\Gamma\left(\frac{5}{z}\right)}{z-5} \right]^{\frac{q+1}{2}} 
\left[ \frac{z-3}{\Gamma\left(\frac{3}{z}\right)}\right]^{\frac{1}{2}} 
[\xi(t)]^{-1-\frac{q}{2}(z-5)} 
\frac{( 1-E_4)^{\frac{q+1}{2}}} 
{(1- E_2)^{\frac{1}{2}}}\;, \label{tuqzg5} 
}\e 
\begin{equation} 
\overline{U}(L,t) = \frac{1}{2\pi} \sqrt{ 
\frac{(z-3)\Gamma\left(\frac{5}{z} \right)} 
{(z-5)\Gamma\left(\frac{3}{z} \right)}}\; 
\left[ \xi(t)\right]^{-1} 
\sqrt{\frac{1-E_4}{1- E_2}}\;, \label{tu0zg5} 
\end{equation} 
and 
\begin{equation} 
 \overline{K}(L,t)=\sqrt{\frac{D\Gamma\left(\frac{5}{z} \right)}{2\nu(z-5)}} 
\left[ \xi(t)\right]^{\frac{z-5}{2}} \sqrt{1-E_4}\; 
 \label{tkzg5} 
 \end{equation} 
and therefore  the time-behaviour is a clean power-law: 
$U_q(L,t)$ {\em decays} as $\sim t^{-[2+q(z-5)]/2z}$, $\overline{U}(L,t) \sim  
t^{-1/z}$, and $\overline{K}(L,t)$ diverges as $\sim t^{(z-5)/2z}$, for $L/\xi 
\gg 1$. 
 
{\em ii)} $z=5$. In this case $\phi_4$ takes the form (\ref{phic2}) but 
$\phi_2$ is still given by (\ref{phic1}). The quantities of interest 
become: 
\bb{ 
U_q(L,t) \simeq \frac{\Gamma\left( \frac{q}{2}+1\right)}{2\pi} 
\left( \frac{2D}{\pi \nu}\right)^{\frac{q}{2}}  
\sqrt{\frac{2}{\Gamma\left( \frac{3}{5}\right)}}\;\xi^{-1} 
\left[ \ln{\left( \frac{\pi \xi}{a}\right)}\right]^{\frac{q+1}{2}} 
\frac{ \left\{ 1+ \left[ \ln{\left( \frac{\pi \xi}{a}\right)}\right]^{-1} 
\left(\frac{{\cal C}}{5}+F_4 \right) \right\}^{\frac{q+1}{2}} 
}{\sqrt{1-E_2}}\;, \label{tuqze5} 
}\e 
\bb{ 
\overline{U}(L,t) \simeq  
\frac{1}{2\pi} \sqrt{\frac{2}{\Gamma\left( \frac{3}{5}\right)}}\; 
\xi^{-1} \sqrt{ \ln{\left( \frac{\pi \xi}{a}\right)}}\; 
\sqrt{\frac{1+ \left[ \ln{\left( \frac{\pi \xi}{a}\right)}\right]^{-1} 
\left(\frac{{\cal C}}{5}+F_4 \right)} 
{\sqrt{1-E_2}}}\;,\label{tu0ze5} 
}\e 
and  
\begin{equation} 
 \overline{K}(L,t)=\sqrt{\frac{D}{2\nu}} 
\sqrt{ \ln{\left( \frac{\pi \xi}{a}\right)}}\; 
\sqrt{1+ \left[ \ln{\left( \frac{\pi \xi}{a}\right)}\right]^{-1} 
\left(\frac{{\cal C}}{5}+F_4 \right)} 
\; \label{tkze5} 
 \end{equation} 
One can observe that the leading temporal behaviour has logarithmic 
component due to the borderline situation: $U_q(L,t)$ {\em decays} as 
$\sim t^{-1/5}(\ln{t})^{(q+1)/2}$, $\overline{U}(L,t) \sim  
t^{-1/5}(\ln{t})^{1/2}$, and $\overline{K}(L,t)$ {\em diverges} as $\sim  
(\ln{t})^{1/2}$. 
 
{\em iii)} $3<z<5$.  
\bb{ 
U_q(L,t) \simeq \frac{\Gamma\left( \frac{q}{2}+1\right)}{2\pi} 
\left( \frac{2D}{\pi \nu}\right)^{\frac{q}{2}}  
\sqrt{\frac{z-3}{\Gamma\left( \frac{3}{z}\right)}} 
(5-z)^{-\frac{q+1}{2}} \left(  
\frac{\pi}{a}\right)^{\frac{q+1}{2}(5-z)} 
\xi^{-\frac{z-3}{2}} \; 
\frac{ \left[ 1-\Gamma\left( \frac{5}{z}\right) 
\left( \frac{a}{\pi \xi}\right)^{5-z} (1-E_4) \right]^ 
{\frac{q+1}{2}}}{\sqrt{1-E_2}}\;, \label{tuq3lzl5} 
}\e 
\bb{ 
\overline{U}(L,t) \simeq  
\frac{1}{2\pi} \sqrt{\frac{z-3}{(5-z) 
\Gamma\left( \frac{3}{z}\right)}}\; 
\left( \frac{\pi }{a}\right)^{\frac{5-z}{2}} 
\xi^{-\frac{z-3}{2}} \; 
\sqrt{\frac{1-\Gamma\left( \frac{5}{z}\right) 
\left( \frac{a}{\pi \xi}\right)^{5-z} (1-E_4) } 
{1-E_2}} 
\;,\label{tu03lzl5} 
}\e 
\bb{ 
\overline{K}(L,t) \simeq 
\sqrt{\frac{D}{2\nu (5-z)}} \left(  
\frac{\pi }{a} 
\right)^{\frac{5-z}{2}}  
\sqrt{1-\Gamma\left( \frac{5}{z}\right) 
\left( \frac{a}{\pi \xi}\right)^{5-z} (1-E_4) } \label{tk3lzl5} 
}\e 

\begin{figure}
\hspace*{3.5cm}\epsfxsize=4 in  
\epsfbox{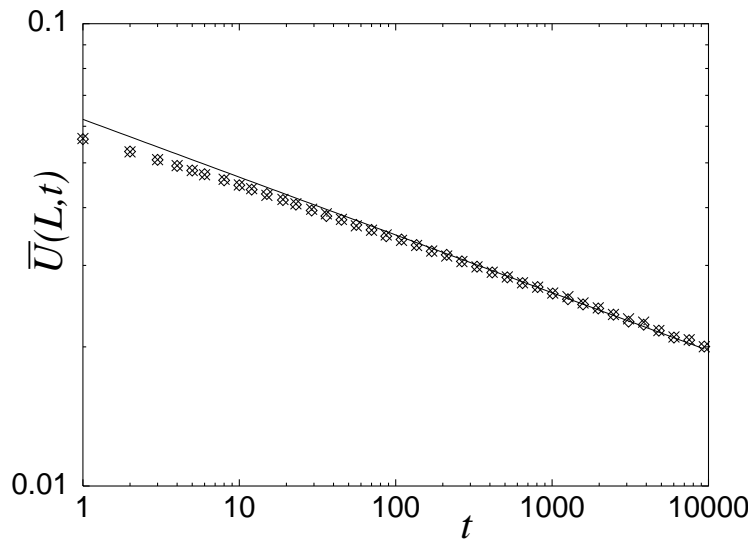} 
\vspace*{0.5cm} 
\caption{Density of minima for the larger curvature model as a function 
of time (the nr of deposited layers), for two system sizes, $L=100$ 
(diamonds) and $L=120$ (crosses). The
straight line corresponds to the behavior $t^{-1/8}$.} 
\end{figure} 

An important conclusion that can be drawn from these 
expressions is that below $z=5$, the leading time-dependence 
of the partition function $U_q(L,t)$  becomes {\em independent} 
of the inverse temperature $q$ and it presents a clean power-law 
decay $\sim t^{-(z-3)/2z}$ which is the same also for  
$\overline{U}(L,t)$.  In particular, for $z=4$ this means a $t^{-1/8}$
decay which is very well verified by the larger curvature model from
Section III.C, see Figure 5. 
Also notice from Eq. (\ref{tu03lzl5}) that the leading term is system
size independent. And indeed, this property is also in a very good agreement
with the numerics on the larger curvature model from Figure 5,
where the two data sets for $L=100$ and $L=120$ practically coincide.

Since the mean curvature depends on $\phi_4$, only, for all 
cases below $z=5$ the dependence is given by the same 
formula (\ref{tk3lzl5}) (just need to replace the corresponding  
value for  $z$).

{\em iv)} $z=3$. This is another borderline situation, 
the corresponding expressions are found easily: 
\bb{ 
U_q(L,t) \simeq \frac{\Gamma\left( \frac{q}{2}+1\right)} 
{2\sqrt{2}\pi} \left( \frac{2D}{\pi \nu}\right)^{\frac{q}{2}}  
\left( \frac{\pi}{a}\right)^{q+1}  
\left[ \ln{\left( \frac{\pi \xi}{a}\right)}\right]^{-\frac{1}{2}} 
\frac{ \left[ 
1-\Gamma\left( \frac{5}{3}\right) 
\left( \frac{a}{\pi \xi}\right)^2 (1-E_4)\right]^{\frac{q+1}{2}}} 
{ \sqrt{ 1+ \left[ \ln{\left( \frac{\pi \xi}{a}\right)}\right]^{-1} 
\left(\frac{{\cal C}}{3}+F_2 \right) }}\;, \label{tuqze3} 
}\e 
\bb{ 
\overline{U}(L,t) \simeq \frac{1}{2\sqrt{2}a}  
\left[ \ln{\left( \frac{\pi \xi}{a}\right)}\right]^{-\frac{1}{2}} 
\sqrt{\frac{1-\Gamma\left( \frac{5}{3}\right) 
\left( \frac{a}{\pi \xi}\right)^2 (1-E_4)} 
{1+ \left[ \ln{\left( \frac{\pi \xi}{a}\right)}\right]^{-1} 
\left(\frac{{\cal C}}{3}+F_2 \right)}}\; \label{tu0ze3} 
}\e 
and the leading time dependences are: $U_q(L,t) \sim 
(\ln{t})^{-1/2}$, $\overline{U}(L,t)  
\sim (\ln{t})^{-1/2}$. 
 
{\em v)} $1<z<3$. 
\bb{ 
U_q(L,t) \simeq \frac{\Gamma\left( \frac{q}{2}+1\right)} 
{2\sqrt{2}\pi} \left( \frac{2D}{\pi \nu}\right)^{\frac{q}{2}}  
\sqrt{\frac{3-z}{(5-z)^{q+1}}} \left(  
\frac{\pi}{a}\right)^{1+\frac{q}{2}(5-z)} 
\frac{ \left[ 1-\Gamma\left( \frac{5}{z}\right) 
\left( \frac{a}{\pi \xi}\right)^{5-z} (1-E_4) \right]^{\frac{q+1}{2}} } 
{\sqrt{1-\Gamma\left( \frac{3}{z}\right) 
\left(\frac{a}{\pi \xi}\right)^{3-z} (1-E_2)}}\;,\label{tuqzl3} 
}\e 
\bb{ 
\overline{U}(L,t) \simeq  
\frac{1}{2a} \sqrt{\frac{3-z}{(5-z)}}\; 
\sqrt{\frac{1-\Gamma\left( \frac{5}{z}\right) 
\left( \frac{a}{\pi \xi}\right)^{5-z} (1-E_4) } 
{1-\Gamma\left( \frac{3}{z}\right) 
\left(\frac{a}{\pi \xi}\right)^{3-z} (1-E_2)}} 
\;,\label{tu0zl3} 
}\e 
In this case the partition function and the density of minima 
all converge to a constant which in leading order is independent 
of the system size. The density of minima was shown in Section 
II to have this property in the steady-state. Here we see not only that 
but also the fact that {\em all} $q$-moments show the same behavior, and 
even more, the time behavior before reaching the steady-state 
constant is not a clean power-law, but rather a decaying correction in the 
approach to this constant. The leading term in the temporal correction 
is of $\sim t^{-(3-z)/z}$ and the next-to-leading has  
$\sim t^{-(5-z)/z}$.

\section{Conclusions and outlook}

In summary, based on the analytical results presented, a 
short wavelength based analysis of interface fluctuations can provide 
us with novel type of information and give an alternative description 
of surface morphologies. This analysis gives a more detailed characterization 
and can be used to distinguish interfaces that are `fuzzy' from those 
that locally appear to be smooth, and the central quantities, the 
extremal-point densities are numerically and analytically accessible. 
 The partition function-like formalism enables us to access a wide range 
of $q$-momenta of the local curvatures distribution. In the case 
of the stochastic evolution equations studied we could exactly relate these 
$q$-momenta to the structure function of the process via the 
simple quantities $\phi_2$ and $\phi_4$. The wide spectrum of results 
accessed through this technique shows the richness of short wavelength 
physics. This physics is there, and the long wavelength approach just  
simply cannot reproduce it, but instead may suggest an oversimplified 
picture of the reality. For example, the MPEU model has been shown to belong 
in the steady state to the EW universality class, however, 
{\em it cannot be described exactly} by the EW  equation in {\em all} 
respects, not even in the steady-state! For example, the utilization (or 
density of minima) of the MPEU model is 0.24641 which for the  
EW model on a lattice is 0.25. Also, if one just simply looks at the  
steady-state configuration, one observes high {\em skewness} 
for the MPEU model  \cite{KTNR},
whereas the EW is completely up-down symmetric. 
This can also be
shown  by comparing the calculated two-slope correlators. 
For a number of models that belong to the  KPZ equation 
universality class, this broken-symmetry property vs. the EW case has been
extensively  investigated by Neergard and den Nijs \cite{NN}. 
The difference on the short wavelength 
scale between two models that otherwise belong to the same universality 
class lies in the existence of irrelevant operators (in the RG sense).  
Although these operators do not change universal properties, 
the quantities associated with them may be of very practical interest.  
The parallel computing example shows that the fundamental 
question of algorithmic scalability is answered  based  on the 
fact that the simulated time horizon in the steady state belongs to the EW 
universality class, thus it has a {\em finite} density of local minima. 
The actual value of the density of local minima in the thermodynamic limit,  
however, strongly depends on the details of the microscopics, which in  
principle can be described in terms of irrelevant operators \cite{NN}. 
 
The extremal-point densities introduced in the present paper may actually 
have a broader application than stochastic surface fluctuations.  
The main geometrical characterization of fractal curves is based on the 
construction of their Haussdorff-Besikovich dimension, or the 
`box-counting' dimension: one covers the set with small boxes of linear size 
$\epsilon$ and then track the divergence of the number of 
boxes needed to cover 
in a minimal way the whole set as $\epsilon$ is lowered to zero. For example, 
a smooth line in the plane has a dimension of unity, but the Weierstrass curve 
of (\ref{weier}) has a dimension of $\ln{b}/\ln{a}$ (for $b>a$). The actual
length of  a fractal curve whose dimension is larger than unity will diverge
when $\epsilon  \to 0^+$. The total length at a given resolution $\epsilon$ is
a {\em global}  property of the fractal, it does not tell us about the way  `it
curves'. The  novel measure we propose in (\ref{minima}) is meant to
characterize the  distribution of a local property of the curve, its
{\em bending} which in turn is a measure of the curve's wrinkledness. For
simplicity we  formulated it for functions, i.e., for curves which are
single-valued in a  certain direction. This can be remedied and generalized by
introducing a  parametrization $\gamma \in [0,1]$ of the curve, and then 
plotting the local {\em curvature} vs. this parameter $K(\gamma)$. The  plot
will be a single valued function on which now (\ref{minima}) is  easily
defined.   

Other desirable extensions of the present technique are: 1) to include a 
statistical description of the degeneracies of higher order, and 2) to repeat 
the analysis for higher (such as $d=2$) substrate dimensions. The latter 
is promising an even richer spectrum of novelties, since in higher dimensions 
there is a plethora of singular points ($\nabla f ={\bf 0} $) which are 
classified by the eigenvalues of the Hessian matrix of the function in the 
singular point. Deciphering the statistical behaviour of these various 
singularities for randomly evolving surfaces is an interesting challenge. 
The studies performed by Kondev and Henley \cite{Kondev} on the distribution of
contours on random Gaussian surfaces should come to a good aid in
achieving this goal.
In particular we may find the method developed here useful in studying the
spin-glass ground state, and the spin-glass transition problem. 
And at last but not the least, we invite the reader to consider instead  of the
Langevin equations studied here, noisy wave equations, with  a second
derivative of the time component, or other stochastic evolution  equations.

\section*{Acknowledgements} 
 
We thank S. Benczik, M.A. Novotny, P.A. Rikvold, 
Z. R\'acz, B. Schmittmann, T. T\'el, E.D. 
Williams, and I. \v{Z}uti\'{c} for stimulating  discussions. This work was 
supported by NSF-MRSEC at University of Maryland, 
by DOE through SCRI-FSU, and 
by NSF-DMR-9871455.

\appendix 
\section{$\langle\Theta(-x_1)\Theta(x_2)\rangle$ for general coupled Gaussian  
variables}  
The expression we derive in this appendix, despite its simplicity, is probably 
the most important one concerning the extremal-point densities of  
one-dimensional Gaussian interfaces on a lattice. If the correlation matrix 
for two possibly coupled Gaussian variables is given by 
\begin{eqnarray} 
\langle x_{1}^{2}\rangle  = \langle x_{1}^{2}\rangle & = & d >0 \\ 
\langle x_{1}x_{2}\rangle & = & c \nonumber 
\end{eqnarray} 
then the distribution follows as 
\begin{equation} 
P(x_{1},x_{2}) = \frac{1}{2\pi\sqrt{\cal D}} 
\exp\left\{ -\frac{1}{2{\cal D}} \left(d x_{1}^{2}+d x_{2}^{2} 
- 2cx_{1}x_{2}\right) \right\} = \frac{1}{2\pi\sqrt{\cal D}} 
\exp\left\{ -\frac{d}{2{\cal D}} \left(x_{1}^{2}+x_{2}^{2}  
- 2\frac{c}{d}x_{1}x_{2}\right) \right\} \;, 
\end{equation} 
where ${\cal D}\equiv d^2 - c^2>0$. We aim to find the average of the  
stochastic variable $u=\Theta(-x_1)\Theta(x_2)$: 
\begin{equation} 
\langle u\rangle = \langle\Theta(-x_1)\Theta(x_2)\rangle =  
\int_{-\infty}^{\infty}\int_{-\infty}^{\infty}d\!x_{1} d\!x_{2} \, 
\Theta(-x_1)\Theta(x_2) P(x_{1},x_{2}) 
\end{equation} 
which is simply the total weight of the density $P(x_{1},x_{2})$ in the 
$x_1 < 0$, $x_2 > 0$ quadrant. If $c = 0$, the density is 
isotropic, and $\langle u\rangle=1/4$. In the general case it is 
convenient to find a new set of basis vectors, where the probability density 
is isotropic (of course the shape of the original quadrant will  
transform accordingly). Introducing the following linear transformation 
\begin{eqnarray} 
x_{1} & = & \sqrt{\frac{{\cal D}}{2}}\left(  
\frac{y_1}{\sqrt{d+c}} + \frac{y_2}{\sqrt{d-c}} \right) \\ 
x_{2} & = & \sqrt{\frac{{\cal D}}{2}}\left(  
-\frac{y_1}{\sqrt{d+c}} + \frac{y_2}{\sqrt{d-c}} \right) \;, \nonumber 
\end{eqnarray} 
and exploiting that $\Theta(\lambda x)=\Theta(x)$ for $\lambda > 0$ 
we have  
\begin{equation} 
\langle u\rangle = 
\int_{-\infty}^{\infty}\int_{-\infty}^{\infty}d\!y_{1} d\!y_{2}\, 
\Theta\left( -\frac{y_1}{\sqrt{d+c}} - \frac{y_2}{\sqrt{d-c}} \right)   
\Theta\left( -\frac{y_1}{\sqrt{d+c}} + \frac{y_2}{\sqrt{d-c}} \right) 
\frac{1}{2\pi}\exp\left\{-\frac{1}{2}(y_1^2 + y_2^2)\right\} 
\end{equation} 
Now the probability density for the new variables, $y_1,y_2$, is isotropic, 
and $\langle u\rangle=\theta/(2\pi)$, 
where $\theta$ is the angle enclosed by the following two unit vectors: 
\begin{equation} 
{\bf v}_1 = \frac{1}{\sqrt{2d}} \left(\begin{array}{r} 
-\sqrt{d+c} \\ \sqrt{d-c} \end{array}\right) \;,\;\;\; 
{\bf v}_2 = \frac{1}{\sqrt{2d}} \left(\begin{array}{r} 
-\sqrt{d+c} \\ -\sqrt{d-c} \end{array}\right) \;. 
\end{equation} 
From their dot product one obtains  
\begin{equation} 
\cos(\theta) = \frac{{\bf v}_1\cdot{\bf v}_2}{|{\bf v}_1||{\bf v}_2|} = 
\frac{c}{d}. 
\end{equation} 
and, thus, for $\langle u\rangle$: 
\begin{equation} 
\langle u\rangle = \frac{1}{2\pi}\arccos\left(\frac{c}{d}\right) \;. 
\end{equation}

\section{Poisson summation formulas} 
 
 In this Appendix we recall the well-known Poisson summation formula 
and adapt it  for functions with finite support in $\R$. In the theory 
of generalized functions \cite{Jones} the following identity is proven: 
\begin{equation} 
\sum_{m=-\infty}^{\infty} \delta(x-m) = \sum_{m=-\infty}^{\infty} 
e^{2\pi i m x} \label{identity} 
\end{equation} 
Let $f: [\alpha,\beta] \to \R$ be a continuous function with continuous 
derivative on the interval $[\alpha,\beta]$. Multiply Eq. (\ref{identity}) on 
both sides with $f(x)$, then integrate both sides  from $\alpha$ to $\beta$. In 
the evaluation of the left hand side we have to pay attention to whether any, 
or both the numbers $\alpha$ and $\beta$ are integers, or non-integers. In the 
integer case the contribution of the end-point is calculated via the identity: 
\begin{equation} 
\int\limits_{n}^{n+r} 
dx\;\delta(x-n)\;f(x)=\frac{1}{2}\;f(n)\;,\;\;\;\forall\;r>0  \label{ident2} 
\end{equation} 
Assuming that $f$ is absolutely integrable if $\beta=\infty$, and choosing 
$\alpha=0$, the classical Poisson summation formula is obtained: 
\begin{equation} 
\sum_{n=0}^{\infty} f(n)=\frac{1}{2}f(0)+\int\limits_{0}^{\infty} dx\;f(x)+ 
2\sum_{m=1}^{\infty}  
\int\limits_{0}^{\infty} dx\;f(x)\;\cos(2\pi m x) \label{CP} 
\end{equation} 
Let us write also explicitely out the case when both $\alpha$ and $\beta$ are 
integers: \begin{equation} 
\sum_{n=\alpha}^{\beta}f(n) = \frac{1}{2}[f(\alpha) + f(\beta)]+ 
\int\limits_{\alpha}^{\beta}dx\;f(x) +2\sum_{m=1}^{\infty}  
\int\limits_{\alpha}^{\beta} dx\;f(x)\;\cos(2\pi m x)\;, 
\;\;\;\mbox{when}\;\;\;\alpha,\beta \in \Z \label{ZP} 
\end{equation} 
Next  we apply these equations to give an exact closed expression for 
the slope correlation function for {\em finite} $L$ [eq. (\ref{C_L})]: 
\begin{equation} 
C^{\phi}_L(l)= 
\frac{D}{L}\sum_{n=1}^{L-1} \frac{e^{i\left( \frac{2\pi n}{L}\right) 
l}}{\nu+2\kappa \left[ 1-\cos\left( \frac{2\pi n}{L}\right) \right]}  
\end{equation} 
where $l \in \{0,1,2,..,L-1 \}$, $\nu,\kappa \in \R^+$. Let us denote 
\begin{equation} 
a=\frac{2 \kappa}{\nu+2\kappa}\;. 
\end{equation} 
We have $|a| <1$, and 
\begin{equation} 
C^{\phi}_L(l)=\frac{Da}{2\kappa L}\sum_{n=1}^{L-1} \frac{e^{i\left( \frac{2\pi 
n}{L}\right) l}}{1-a\cos\left( \frac{2\pi n}{L}\right)}  
\end{equation} 
In order to apply the Poisson summation formula (\ref{ZP}), we introduce 
the function: 
\begin{equation} 
f(x)=\frac{a}{2\kappa L}\sum_{n=1}^{L-1} \frac{e^{i\left( \frac{2\pi 
x}{L}\right) l}}{1-a\cos\left( \frac{2\pi x}{L}\right)} \;, 
\;\;\;1\leq x \leq L-1 
\end{equation} 
and identify in (\ref{ZP}) $\alpha \equiv 1$ and $\beta \equiv L-1$.  
The non-integral terms of (\ref{ZP}) give: 
\begin{equation} 
\frac{1}{2}[f(1) + f(L-1)] = \frac{a}{2 \kappa L}\;\frac{\cos\left( 
\frac{2\pi}{L}l\right)}{1- a\cos\left( 
\frac{2\pi}{L}l\right)} \label{first} 
\end{equation} 
The next term becomes: 
\begin{equation} 
\int\limits_{1}^{L-1}dx\;f(x)=\frac{a}{2\kappa \sqrt{1-a^2}}\; 
\left( \frac{1-\sqrt{1-a^2}}{a}\right)^l-\frac{a}{2 \pi \kappa} 
\int\limits_{0}^{2 \pi / L}dx\; \frac{\cos{xl}}{1-a\cos{x}} 
\label{second}  
\end{equation} 
where during the evaluation of the integral we made a simple 
change of variables and used a  well-known integral from  
random walk theory \cite{Hughes}, \cite{RG}: 
\begin{equation} 
\int\limits_{-\pi}^{\pi}dx\;\frac{e^{ixl}}{1-a\cos{x}}= 
\frac{2 \pi }{ \sqrt{1-a^2}}\; 
\left(\frac{1-\sqrt{1-a^2}}{a}\right)^l\;,\;\;\;l\geq 0 \label{rwalk}  
\end{equation} 
The sum over the integrals in (\ref{ZP}) can  also be 
evaluated, and one obtains: 
\begin{equation} 
2\sum_{n=1}^{\infty}  
\int\limits_{1}^{L-1} dx\;f(x)\cos(2\pi n x) =  
\frac{a \left(b^l + b^{-l} \right)}{2 \kappa \sqrt{1-a^2}}\;\frac{b^L}{1-b^L} 
-\frac{a}{2 \pi \kappa} \sum_{n=1}^{\infty} \int\limits_{-2 \pi /L}^{2 \pi /L} 
dx\;\cos(nLx)\;\frac{e^{ilx}}{1-a\cos{x}} \label{akarmi} 
\end{equation} 
where 
\begin{equation} 
b=\frac{1-\sqrt{1-a^2}}{a}\;,\;\;\;\mbox{and}\;\;\;|b|<1 
\end{equation} 
To compute the sum on the rhs of (\ref{akarmi}) we recall another 
identity from the theory of generalized functions (see Ref. \cite{Jones}, 
page 155): 
\begin{equation} 
\sum_{n=1}^{\infty} e^{i n x}  = \pi \sum_{m=-\infty}^{\infty} 
\delta(x-2m\pi) +  
\frac{i}{2} ctg\left( \frac{x}{2}\right)-\frac{1}{2} \label{identity'}  
\end{equation} 
Combining (\ref{identity'}) and identity (\ref{identity}), one obtains: 
\begin{equation} 
\sum_{n=1}^{\infty} \cos(nx) = \pi \sum_{m=-\infty}^{\infty} 
\delta(x-2m\pi) + \frac{1}{2} \label{identity''}  
\end{equation} 
Peforming the sum over $n$ directly in the rhs of (\ref{akarmi}) via 
(\ref{identity''}), yields: 
\begin{equation} 
-\frac{a}{2 \pi \kappa} \sum_{n=1}^{\infty} \int\limits_{-2 \pi /L}^{2 \pi /L} 
dx\;\frac{\cos(nLx)e^{ilx}}{1-a\cos{x}} =  
-\frac{a}{2 \kappa L}\sum_{m=-\infty}^{\infty} \int\limits_{-2\pi}^{2\pi} 
dy\;\frac{e^{ily}\delta (y-2m\pi) }{1-a\cos{y}}  + \frac{a}{2\pi \kappa} 
\int\limits_{0}^{2\pi /L} dx\;\frac{\cos{lx}}{1-a\cos{x}} \label{tt} 
\end{equation} 
Only $m=\pm 1, 0$ contribute in (\ref{tt}). With the help of  
(\ref{ident2}): 
\begin{equation} 
-\frac{a}{2 \pi \kappa} \sum_{n=1}^{\infty} \int\limits_{-2 \pi /L}^{2 \pi /L} 
dx\;\frac{\cos(nLx)e^{ilx}}{1-a\cos{x}} =  
-\frac{a}{2 \kappa L}\left\{\frac{1}{1-a}+ 
\frac{\cos\left(\frac{2\pi}{L}l\right)}{1-a\cos\left(\frac{2\pi}{L}\right)} 
\right\} + \frac{a}{2\pi \kappa} 
\int\limits_{0}^{2\pi /L} dx\;\frac{\cos{lx}}{1-a\cos{x}}\label{ttt} 
\end{equation} 
Using (\ref{ttt}) in (\ref{akarmi}), we can add the result to the 
rest of the contributions (\ref{first}) and (\ref{second}) to obtain 
the final expression [eq. (\ref{C_phi})] after the cancellations.


\end{document}